\documentclass[aps,superscriptaddress,prc,twocolumn,nofootinbib]{revtex4}

\usepackage[]{graphicx}
\usepackage{color}
\usepackage{hyperref}
\usepackage{amsmath,amssymb,amsfonts}

\usepackage{slashed}

\usepackage{epstopdf}
\usepackage{subfigure}
\usepackage{epsfig}
\usepackage[utf8]{inputenc}

\begin{document}

\title{Heavy and light flavor jet quenching in different collision systems at the LHC energies}

\author{Yu-Fei Liu}
\affiliation{Institute of Particle Physics and Key Laboratory of Quark and Lepton Physics (MOE), Central China Normal University, Wuhan, Hubei, 430079, China}

\author{Wen-Jing Xing}
\email{wenjing.xing@mails.ccnu.edu.cn}
\affiliation{Institute of Particle Physics and Key Laboratory of Quark and Lepton Physics (MOE), Central China Normal University, Wuhan, Hubei, 430079, China}

\author{Xiang-Yu Wu}
\affiliation{Institute of Particle Physics and Key Laboratory of Quark and Lepton Physics (MOE), Central China Normal University, Wuhan, Hubei, 430079, China}

\author{Guang-You Qin}
\email{guangyou.qin@ccnu.edu.cn}
\affiliation{Institute of Particle Physics and Key Laboratory of Quark and Lepton Physics (MOE), Central China Normal University, Wuhan, Hubei, 430079, China}

\author{Shanshan Cao}
\email{shanshan.cao@sdu.edu.cn}
\affiliation{Institute of Frontier and Interdisciplinary Science, Shandong University, Qingdao, Shandong 266237, China}

\author{Hongxi Xing}
\email{hxing@m.scnu.edu.cn}
\affiliation{Guangdong Provincial Key Laboratory of Nuclear Science, Institute of Quantum Matter, South China Normal University, Guangzhou 510006, China}
\affiliation{Guangdong-Hong Kong Joint Laboratory of Quantum Matter, Southern Nuclear Science Computing Center, South China Normal University, Guangzhou 510006, China}

\date{\today}
\begin{abstract}

Recent experiments have observed large anisotropic collective flows in high multiplicity proton-lead collisions at the Large Hadron Collider (LHC), which indicates the possible formation of mini quark-gluon plasma (QGP) in small collision systems.
However, no jet quenching has been confirmed in such small systems so far.
To understand this intriguing result, the system size scan experiments have been proposed to bridge the gap between large and small systems.
In this work, we perform a systematic study on both heavy and light flavor jet quenching in different collision systems at the LHC energies.
Using our state-of-the-art jet quenching model, which combines the next-to-leading-order perturbative QCD framework, a linear Boltzmann transport model and the (3+1)-dimensional viscous hydrodynamics simulation, we provide a good description of nuclear modification factor $R_{\rm AA}$ for charged hadrons and $D$ mesons in central and mid-central Pb+Pb and Xe+Xe collisions measured by CMS collaboration.
We further predict the transverse momentum and centrality dependences of $R_{AA}$ for charged hadrons, $D$ and $B$ mesons in Pb+Pb, Xe+Xe, Ar+Ar and O+O collisions at the LHC energies.
Our numerical results show a clear system size dependence for both light and heavy flavor hadron $R_{AA}$ across different collision systems.
Sizable jet quenching effect is obtained for both heavy and light flavor hadrons in central O+O collisions at the LHC energies.
Our study provides a significant bridge for jet quenching from large to small systems, and should be helpful for finding  the smallest QGP droplet and the disappearance of QGP in relativistic nuclear collisions.

\end{abstract}
\maketitle

\section{Introduction}

Jet quenching and anisotropic flows are the two most important signatures of the quark-gluon plasma (QGP) created in high-energy nucleus-nucleus collisions performed at the Relativistic Heavy-Ion Collider (RHIC) and the Large Hadron Collider (LHC).
At low transverse momentum ($p_T$) region, the produced final-state hadrons exhibit large anisotropies in the azimuthal angle distribution. Such anisotropy can be successfully explained by relativistic hydrodynamics simulation \cite{Romatschke:2017ejr, Rischke:1995ir, Heinz:2013th, Gale:2013da, Huovinen:2013wma}, which suggests that the produced hot and dense nuclear matter is a strongly-interacting QGP.
The strong interaction among QGP constituents can convert initial-state geometric anisotropy into final-state anisotropy of hadron momentum distributions \cite{Alver:2010gr, Petersen:2010cw, Qin:2010pf, Staig:2010pn, Teaney:2010vd, Schenke:2010rr, Ma:2010dv, Qiu:2011iv}.
At high $p_T$ region, the final-state hadron yields are strongly suppressed compared to the binary-scaled nucleon-nucleon collisions. Such phenomenon can be well explained by the energy loss of hard jet partons during their propagation through the soft QGP medium, and is usually called jet quenching  \cite{Wang:1991xy, Qin:2015srf, Blaizot:2015lma, Majumder:2010qh, Gyulassy:2003mc, Cao:2020wlm}.
The interaction between jets and medium may also lead to the modification of jet-related correlations \cite{Qin:2009bk, Chen:2016vem, Chen:2016cof, Chen:2017zte, Luo:2018pto, Zhang:2018urd, Kang:2018wrs}, the suppression of full jet yields \cite{Qin:2010mn, Young:2011qx, Dai:2012am, Wang:2013cia, Blaizot:2013hx, Mehtar-Tani:2014yea, Cao:2017qpx, Kang:2017frl, He:2018xjv, Chang:2016gjp, Casalderrey-Solana:2016jvj} and the change of jet structure and substructure \cite{Tachibana:2017syd, KunnawalkamElayavalli:2017hxo, Brewer:2017fqy, Chien:2016led, Milhano:2017nzm, Chang:2019sae}.
Currently, one important task of heavy-ion physics is to quantitatively extract various transport coefficients of the QGP and jet transport parameters using low and high $p_T$ data \cite{Bernhard:2019bmu, Everett:2020xug, Burke:2013yra, Cao:2021keo, Cao:2018ews, Rapp:2018qla}, and ultimately to probe the structure and transport properties of QGP at various energy and length scales.

In recent years, experiments have observed large anisotropic flows in proton-lead collisions at the LHC and deuteron-gold collisions at RHIC \cite{Abelev:2012ola, Aad:2012gla, Chatrchyan:2013nka, Adare:2013piz}. The successfulness of relativistic hydrodynamics in explaining the collectivity in such small systems implies the possible formation of mini QGP \cite{Bozek:2013uha, Bzdak:2013zma, Qin:2013bha, Werner:2013ipa, Bozek:2013ska, Nagle:2013lja, Schenke:2014zha, Zhao:2019ehg, Zhao:2017rgg}.
However, no jet quenching has been confirmed in these small systems so far, e.g., $R_{pA} \approx 1$.
On the other hand, the color glass condensate framework shows that the initial-state effect, i.e., the interaction between partons originated from the nucleon projectile and dense gluons inside the target nucleus before the onset of hydrodynamics evolution, can also generate significant amount of collectivity \cite{Dusling:2012iga, Dusling:2017dqg, Mace:2018vwq, Davy:2018hsl, Zhang:2019dth}.
Nowadays, how to disentangle the contributions from initial and final state effects is still in hot debate.

In order to understand the above intriguing results, the system size scan has been proposed to bridge the gap between large and small collision systems \cite{Citron:2018lsq, Huang:2019tgz}.
By scanning different sizes of collision systems, we may build a smooth transition from large to small collision systems, and help to disentangle the initial and final state contributions.
Along this line, the system size scan predictions for Pb+Pb, Xe+Xe, Ar+Ar, O+O and other collision systems have been performed from relativistic hydrodynamics \cite{Sievert:2019zjr, Lim:2018huo}.
As for hard probes, there have been studies on the nuclear modification factor $R_{AA}$ for Xe+Xe collisions \cite{Zigic:2018ovr, Shi:2019nyp} and O+O collisions \cite{Huss:2020dwe, Huss:2020whe} and the system scan study for  heavy quarks and heavy flavor mesons \cite{Katz:2019qwv}.
However, a systematic study of jet quenching in different collision systems for both light and heavy flavor partons and hadrons is still lacking.
This is the main purpose of our work.

In this work, we perform a systematic study on both heavy and light flavor jet quenching in different collision systems at the LHC energies.
The productions of jet partons and hadrons are calculated within a next-to-leading-order perturbative QCD framework, the evolution of heavy and light flavor jet partons inside the QGP is simulated via a linear Boltzmann transport model, and the space-time profile of the QGP fireball is obtained via a (3+1)-dimensional viscous hydrodynamics simulation.
Using our state-of-the-art jet quenching model \cite{Xing:2019xae}, we obtain a nice description of $R_{\rm AA}$ for charged hadrons and $D$ mesons in central and mid-central Pb+Pb and Xe+Xe collisions measured by CMS collaboration.
Using the same setup, we further compute the $p_T$ and centrality dependences of $R_{AA}$ for charged hadrons, $D$ and $B$ mesons in Pb+Pb, Xe+Xe, Ar+Ar and O+O collisions at the LHC energies.
Our study shows a clear system size dependence for both light and heavy flavor hadron $R_{AA}$ across different collision systems.
This suggests that $R_{pA} \approx 1$ in proton-nucleus collisions is mainly due to the small size of the collision system.
Interestingly, we observe sizable jet quenching effect for both heavy and light flavor hadrons in central O+O collisions at the LHC energies, which can be tested in future experiments.

The paper is organized as follows. In Section II, we briefly present the next-to-leading-order framework for computing the momentum spectra of hard jet partons and high $p_T$ hadron. In Section III, we provide the linear Boltzmann transport model for describing the evolution of jet partons interacting with the soft QGP medium. In Section IV, we discuss the setups of the initial conditions and hydrodynamics simulation for different collision systems. In Section V, we provide the numerical results on  $R_{AA}$ as a function of $p_T$ and centrality for charged hadrons, $D$ mesons, $B$ mesons and heavy flavor decayed electrons in Pb+Pb, Xe+Xe, Ar+Ar and O+O collisions at the LHC energies. Section VI contains our summary.


\section{NLO perturbative QCD framework}

In this work, we use the next-to-leading-order (NLO) perturbative QCD framework developed in Refs.~\cite{Jager:2002xm, Aversa:1988vb} to calculate the production of both light and heavy flavor partons and hadrons at high transverse momentum ($p_T$) region. In this framework, the differential cross section for high $p_T$ hadrons can be written as follows:
\begin{align}
E_h\frac{d\sigma_{pp \to h}}{d^3p_h} &= \frac{1}{\pi S} \sum_{abc} \int \frac{dz_c}{z_c^2} \int \frac{dv}{v(1-v)} \int \frac{dw}{w}
\nonumber\\&
\times  f_a(x_a,\mu)  f_b(x_b,\mu) \frac{d\sigma_{ab\to c}}{dvdw} D_{h/c}(z_c,\mu).
\end{align}
In the above equation, the sum $\sum_{abc}$ is over all possible partonic species.  The integral ranges are set as: $z_c \in [1 - V + VW, 1]$, $v\in [VW/z_c, 1-(1-V)/z_c]$, $w \in [VW/(vz_c), 1]$, where the hadronic variables $S$, $T$, $U$, $V$, $W$ are defined as:
\begin{align}
& V = 1 + \frac{T}{S},\, W = - \frac{U}{S+T},
\nonumber\\
& S = (p_A+p_B)^2,\, T=(p_A - p_h)^2,\, U = (p_B-p_h)^2.
\end{align}
The corresponding partonic variables are defined as:
\begin{align}
& v = 1+ \frac{t}{s},\, w = - \frac{u}{s+t},
\nonumber\\
& s = (p_a+p_b)^2,\, t=(p_a-p_c)^2,\, u=(p_b-p_c)^2.
\end{align}
$f_a(x_a)$ and $f_b(x_b)$ are parton distribution functions (PDFs), and $D_{h/c}$ is fragmentation function (FF).
Neglecting all masses, one has the following relations:
\begin{align}
& x_a = \frac{VW}{vwz_c},\, x_b = \frac{1-V}{(1-v)z_c},
\nonumber\\
& s = x_a x_b S,\, t = \frac{x_a}{z_c} T,\, u = \frac{x_b}{z_c} U.
\end{align}
At one-loop level, the partonic cross section takes the following form:
\begin{align}
\frac{d\sigma_{ab\to c}}{dvdw} = \frac{d\sigma_{ab\to c}^{(0)}}{dv}\delta(1-w) + \frac{\alpha_s(\mu)}{\pi} \frac{d\sigma_{ab\to c}^{(1)}}{dvdw}.
\end{align}
Since the above NLO perturbative QCD framework neglects all masses, it is expected to be applicable in high $p_T$ regime, $p_T \gg m_c, m_b$. In this work, we focus on heavy and light flavor hadron productions at high $p_T$ ($p_T >$~6-8~GeV), where the fragmentation mechanism is also the dominant contribution for hadron production.
In our calculation, the parton distribution functions are taken from CTEQ parameterizations~\cite{Pumplin:2002vw}. The cold nuclear matter (CNM) effects on PDFs in nucleus are taken from EPS09 parameterizations~\cite{Eskola:2009uj}.
The fragmentation functions are taken from Refs.~\cite{Kretzer:2000yf, Kneesch:2007ey, Kniehl:2008zza} for charged hadrons, $D$ mesons, and $B$ mesons, respectively.

\begin{figure*}
\includegraphics[width=0.440\textwidth]{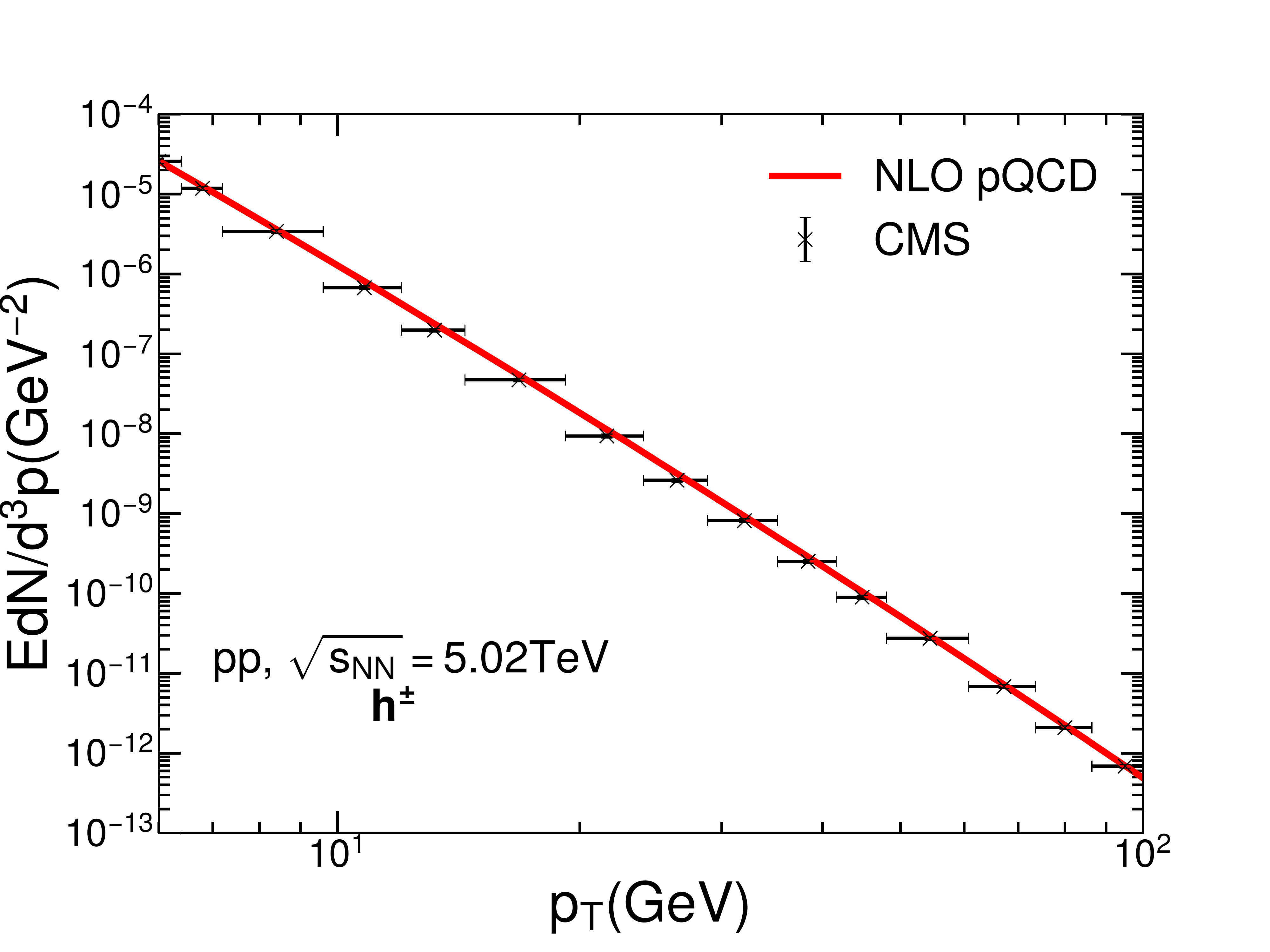}
\includegraphics[width=0.440\textwidth]{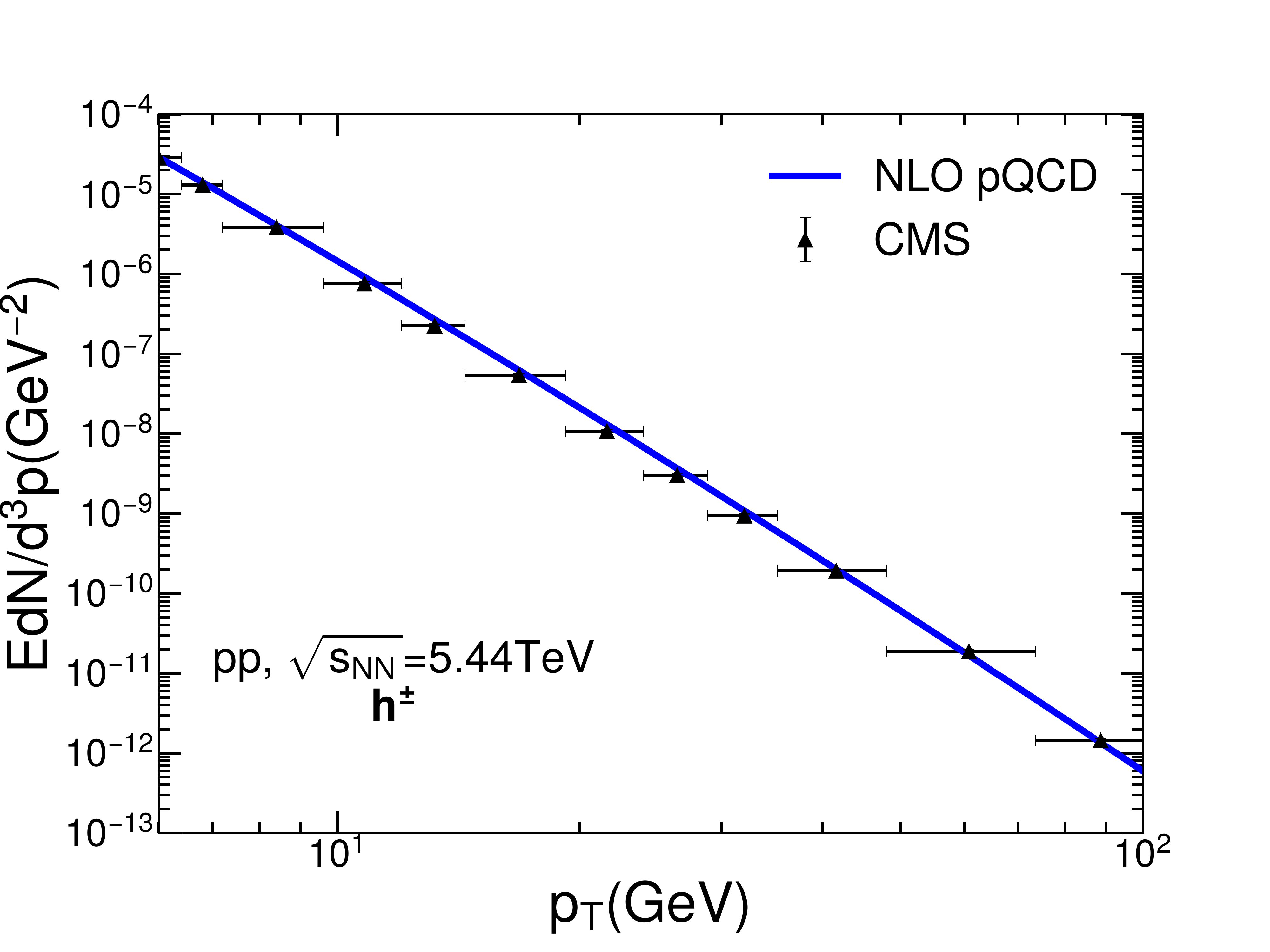}
\includegraphics[width=0.440\textwidth]{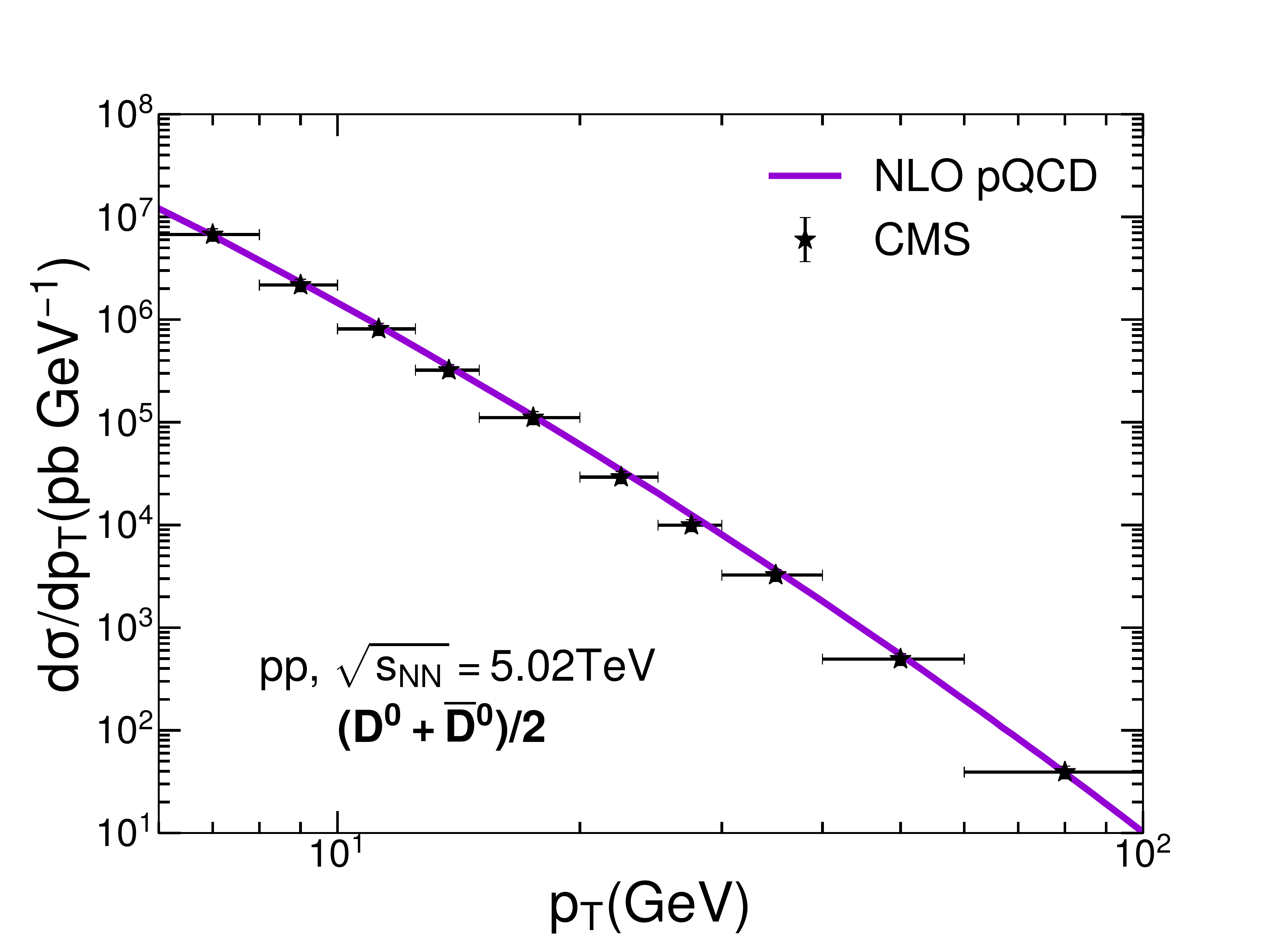}
\includegraphics[width=0.440\textwidth]{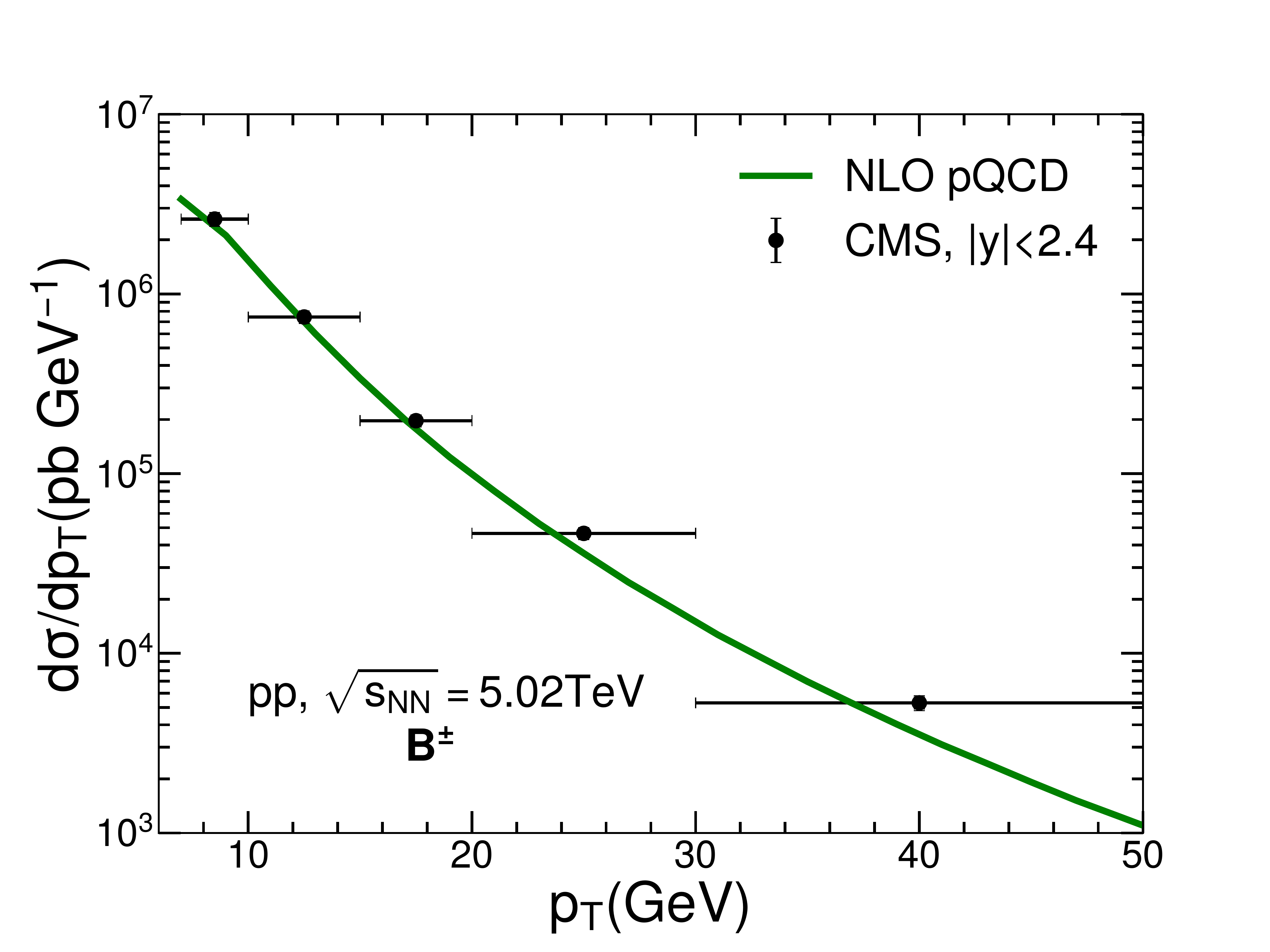}
\caption{The $p_T$ spectra of charged hadrons, $D^0$ mesons, $B$ mesons in p+p collisions from NLO perturbative QCD calculations compared to CMS data~\cite{Khachatryan:2016odn, Sirunyan:2017xss, Sirunyan:2018eqi, Sirunyan:2017oug}.}
\label{dsigma_pp}
\end{figure*}

In Fig.~\ref{dsigma_pp}, we show the transverse momentum spectra for charged hadrons, $D$ mesons and $B$ mesons in p+p collisions. The upper two plots are charged hadron spectra in p+p collisions at $\sqrt{s} = 5.02$~TeV and $\sqrt{s} = 5.44$~TeV, and the lower two plots are $D$ and $B$ meson spectra in p+p collisions at $\sqrt{s} = 5.02$~TeV. In the NLO perturbative QCD calculation, the scales in the above equations are all taken as the transverse momenta of hadrons $\mu = p_T$.
One can see that the NLO perturbative QCD framework can provide a nice description of both heavy and light flavor hadron production at high $p_T$ region in nucleon-nucleon collisions, as measured by CMS collaboration \cite{Khachatryan:2016odn, Sirunyan:2017xss, Sirunyan:2018eqi, Sirunyan:2017oug}.
This serves as the baseline to study the hot QGP medium modification effect in high-energy nucleus-nucleus collisions.

\section{Linear Boltzmann transport model}

We use linear Boltzmann transport (LBT) model to simulate the elastic and inelastic interactions between hard partons and the soft QGP \cite{Wang:2013cia, He:2015pra, Cao:2016gvr, Cao:2017hhk, Xing:2019xae}.
In the LBT model, one numerically uses the Monte-Carlo method to solve the following Boltzmann equation for the propagating hard parton ``1":
\begin{align}
  \label{eq:boltzmann1}
  p_1\cdot\partial f_1(x_1,p_1)=E_1 (C_\mathrm{el} + C_\mathrm{inel}),
\end{align}
where $C_\mathrm{el}$ and $C_\mathrm{inel}$ represent the collision integrals for elastic and inelastic interactions between
the hard parton and the medium constituents.

To simulate the elastic process ($1+2 \rightarrow 3+4$), the key quantity is to calculate the elastic scattering rate:
\begin{align}
\label{eq:gamma0}
\Gamma_{\rm el} (\vec{p}_1)
& = \frac{\gamma_2}{2E_1} \int \frac{d^3 p_2}{(2\pi)^3 2E_2} \int \frac{d^3 p_3}{(2\pi)^3 2E_3}\int \frac{d^3 p_4}{(2\pi)^3 2E_4}
\nonumber\\
& \times f_2(\vec{p}_2)[1\pm f_3(\vec{p}_3)] [1\pm f_4(\vec{p}_4)] S_2(s,t,u)
\nonumber\\
& \times (2\pi)^4 \delta^{(4)} (p_1 + p_2 - p_3 -p_4)|M_{12 \rightarrow 34}|^2.
\end{align}
In the above equation, $\gamma_2$ is the degeneracy factor of parton ``2".
The condition $S_2(s,t,u) = \theta(s\geq 2 \mu_D^2)\theta(t \leq-\mu_D^2)\theta(u \leq-\mu_D^2)$ is imposed to avoid possible divergence at small angle $u, t \rightarrow 0$~\cite{Auvinen:2009qm, He:2015pra},
with the Debye screening mass taken as $\mu_D^2 = 6\pi \alpha_s T_m^2$, with $T_m$ the temperature of the QGP medium.
The $\pm$ sign represents the Pauli block and Bose enhancement effects for the thermal partons.
The factor $(1-f_3)$ is neglected in the calculation due to the much lower density of hard partons in the QGP.
The matrix element $|M_{12 \rightarrow 34}|^2$  is taken from the leading order perturbative QCD calculation~\cite{Combridge:1978kx}.
Since we only consider the leading-order $2 \to 2$ scattering processes, the distribution for the exchanged transverse momentum ($q_\perp$) between hard partons and QGP typically has a hard tail.
To account for the possible contributions from multiple soft scatterings, we impose an effective cutoff $q_\perp < 10 T_m$ for exchanged transverse momentum.
Such cutoff will reduce the parton energy loss, which can be compensated by shifting effective strong coupling $\alpha_{\rm s}$ to a larger value.
In this work, we use $\alpha_\mathrm{s}=0.25$ for the interaction vertex connecting to the thermal partons inside the medium.
For the vertices connecting to jet partons, we set $\alpha_s=4\pi /\left[9\ln (2E_1T_m)/\Lambda^2) \right]$ with $\Lambda=0.2~\text{GeV}$.

To simplify the above elastic scattering rate, we may choose the jet parton momentum $\vec{p}_1$ along the $+z$ direction and the medium parton $\vec{p}_2$ in the $xz$ plane.
Then the elastic scattering rate can be obtained as follows:
\begin{align}
\Gamma_{\rm el} (\vec{p}_1) &=\frac{\gamma_2}{16 E_1 (2 \pi)^4}\int dE_2 d\theta_2 d\theta_4 d\phi_4 \nonumber\\
 &\times f_2(E_2,T_m)(1\pm f_4(E_4,T_m))S_2(s,t,u) \big| {M_\mathrm{12\rightarrow34}} \big|^2
 \nonumber\\
 &\times \frac{p_2 p_4 \sin \theta_2 \sin \theta_4}{E_1+E_2-p_1 \cos \theta_4 -p_2 \cos\theta_{24} },
 \label{eq:gamma1}
\end{align}
where  $\theta_2$ is the polar angle of the momentum $\vec{p}_2$,
$(\theta_4, \phi_4)$ are the polar and azimuthal angles of the momentum $\vec{p}_4$, and $\theta_{24}$ is the angle between $\vec{p}_2$ and $\vec{p}_4$:
\begin{align}
\cos\theta_{24} = \sin \theta_2 \sin \theta_4 \cos \phi_4+ \cos \theta_2 \cos \theta_4.
\end{align}
From energy-momentum conservation, one can find,
\begin{align}
E_4  = \frac{p_1 \cdot p_2 + m_4^2}{E_1 + E_2 - p_1\cos \theta_4 - p_2\cos \theta_{24}}.
\end{align}
Using the above formula, one can calculate the elastic scattering rate $\Gamma_{\rm el}^{(i)}$ for each channel and the total scattering rate $\Gamma_{\rm el} = \sum_{i} \Gamma_{\rm el}^{(i)}$.
If the number of elastic scatterings in a given time step $\Delta t$ follows the Poisson distribution,
then the probability for having elastic scatterings in $\Delta t$ is given by:
\begin{align}
	P_{\rm el} = 1 - e^{-\Gamma_{\rm el} \Delta t}.
\end{align}
For inelastic radiative process, we use the higher-twist formalism for the medium-induced gluon radiation spectrum~\cite{Wang:2001ifa, Zhang:2003wk},
\begin{align}
\frac{dN_g}{dx dk_\perp^2 dt}=&\frac{2 \alpha_s C_A P(x)k_\perp^4}{\pi (k_\perp^2+x^2M^2)^4} \hat{q}(E_1,T_m)\nonumber \\& \times \sin^2 \bigg(\frac{t-t_i}{2\tau_f} \bigg),
\end{align}
where $x$ and $k_\perp$ are the fraction energy and transverse momentum of the emitted gluon,
$M$ is the mass of the parent parton (here we use $M_c = 1.27$~GeV/c and $M_b = 4.19$~GeV/c),
$\alpha_s$ is the strong coupling for the splitting vertex, $P(x)$ is the vacuum splitting function,
$\tau_f=2Ex (1-x)/(k_\perp^2+x^2 M^2)$ is the formation time for gluon radiation,
$(t-t_i)$ denotes accumulative time for a gluon radiation event, i.e., $t_i$ is reset to zero after each radiation, $\hat{q}(E_1,T_m)$ is the jet transport parameter characterizing the transverse momentum broadening rate from elastic collisions and can be calculated as follows:
\begin{align}
\hat{q}&(E_1,T_m)=\frac{\gamma_2}{16 E_1 (2 \pi)^4}\int dE_2 d\theta_2 d\theta_4 d\phi_4 (\vec{p}_{2\perp}-\vec{p}_{4\perp})^2 \nonumber\\
 &\times f_2(E_2,T_m)(1\pm f_4(E_4,T_m))S_2(s,t,u)\big| {M_\mathrm{12\rightarrow34}} \big|^2\nonumber\\ &
 \times \frac{p_2 p_4 \sin \theta_2 \sin \theta_4}{E_1+E_2-p_1 \cos \theta_4 -p_2 \cos\theta_{24} }.
\end{align}
The above formula naturally includes the dependences of $\hat{q}$ on jet energy and medium temperature across
different colliding systems investigated in this work. Using the above gluon radiation formula, one may calculate the average gluon radiation rate in a give time step $\Delta t$ as follows:
\begin{align}
\Gamma_{\rm inel} = \int dx dk_\perp^2 \frac{dN_g}{dx dk_\perp^2 dt},
\end{align}
In this work, we impose a lower cut-off $x_{\rm min}=\mu_D/E$ for the emitted gluon energy to avoid the possible divergence as $x\to 0$.
If the number of radiated gluons follows a Poisson distribution, then the probability for having inelastic scatterings
in a given time step $\Delta t$ is given by:
\begin{align}
P_\mathrm{inel}=1-e^{-\Gamma_{\rm inel}\Delta t}.
\end{align}
To include both elastic and inelastic interactions, the total scattering rate is given by:
\begin{align}
\Gamma_{\rm tot} = \Gamma_{\rm el}+\Gamma_{\rm inel}.
\end{align}
The total scattering probability is then obtained as:
\begin{align}
P_{\rm tot} = 1-e^{-\Gamma_{\rm tot} \Delta t} = P_{\rm el} + P_{\rm inel} - P_{\rm el} P_{\rm inel}.
\end{align}
One can split the total interaction probability into two parts: $P_{\rm el} (1 - P_{\rm inel})$  for pure elastic scatterings
and $P_{\rm inel}$ for inelastic scatterings.


\section{Hydrodynamics simulation and initial conditions}

In this work, we use the (3+1)-dimensional CLVisc~\cite{Pang:2018zzo} hydrodynamics model to simulate
the space-time evolution of QGP created in different collision systems: Pb+Pb collisions at 5.02A~TeV, Xe+Xe collisions at 5.44A~TeV, Ar+Ar collisions at 5.85A~TeV and O+O collisions at 6.5A~TeV.
The initial conditions for hydrodynamics simulation are obtained as follows.
The entropy distribution $S(\tau_0,x,y,\eta_s)$ at the initial proper time $\tau_0$ is factorized into
the entropy density $s(x,y)$ in the transverse plane and the longitudinal envelop function $H(\eta_s)$,
\begin{equation}
S(\tau_0,x,y,\eta_s) = Ks(x,y)H(\eta_s),
\end{equation}
where $K$ is an overall normalization factor which is adjusted to describe the charged hadron spectra in the most central collisions~\cite{Adam:2016ddh, Acharya:2018hhy}. For Ar-Ar and O-O collisions at the LHC energies, where the experimental data on multiplicity are unavailable, we use the empirical power-law formula: ${2}/{\left< N_{\text{part}} \right>} \left<  dN_{\text{ch}}/d\eta\right>|_{|\eta|<0.5}=0.7455 (s_{\rm NN}/{\rm GeV}^2)^{0.1538}$ \cite{Acharya:2018hhy}, which is fitted to the multiplicity data of RHIC Au+Au and LHC Pb+Pb collisions at various colliding energies.
The entropy density $s(x,y)$ in the transverse plane is generated from the TRENTo model\cite{Moreland:2014oya},
in which the positions of nucleons in nucleus are sampled according to the following Woods-Saxon distribution:
\begin{equation}
\rho(r,\theta) = \rho_0 \frac{1+w\frac{r^2}{R^2(\theta)}}{1+\exp\left(\frac{r-R(\theta)}{d}\right)},
\end{equation}
where $\rho_0$ denotes the nuclear density at the center, $d$ is a surface thickness parameter, the radius parameter
\begin{align}
R(\theta)=R_0[1+\beta_2 Y_{20}(\theta) + \beta_4 Y_{40}(\theta)],
\end{align}
with $Y_{nl}(\theta)$ being the spherical harmonic functions, and $(w, \beta_2, \beta_4)$ controlling the deviations from spherical shape.
Table.(\ref{tab:Woods}) shows the parameters used the Woods-Saxon distribution for different collision systems.

\begin{table}[h]
\centering
\begin{tabular}{|c|c|c|c|c|c|c|c|}
\hline
	System & $\sqrt{s_{\rm NN}}$ & $R_0$ (fm) & $d$ (fm) & $\omega$ & $\beta_2$ & $\beta_4$ & $\sigma_{\rm inel}^{\rm NN}$(mb)\\ \hline
	PbPb & 5.02TeV & 6.62 & 0.546 & 0     & 0 & 0 & 70 \\ \hline XeXe & 5.44TeV & 5.4 & 0.59 & 0     & 0.18 & 0 & 70 \\ \hline
	ArAr & 5.85TeV & 3.53 & 0.542 & 0      & 0 & 0 & 71 \\ \hline
	OO & 6.5TeV & 2.608 & 0.513 & -0.051 & 0 & 0 & 72.5 \\ \hline
\end{tabular}
\caption{The parameters in the Woods-Saxon distribution and Glauber model for different collision systems~\cite{Acharya:2018hhy,Sievert:2019zjr}.}
\label{tab:Woods}
\end{table}

Using the Glauber model simulation, one may calculate the nuclear thickness function $T_A(x,y)$ and $T_B(x,y)$
by assigning each participant nucleon a Gaussian smearing function (with width 0.5~fm).
Then the local entropy density $s(x,y)$ is constructed from the thickness function according to the following formula,
\begin{equation}
s(x,y) = \left(\frac{T_A^p + T_B^p}{2}\right)^{\frac{1}{p}}.
\end{equation}
In this work, we choose the parameter $p=0$ which mimics the IP-Plasma or EKRT initial condition models.
For the longitudinal direction, we use the following parameterized form for the envelop function $H(\eta_s)$~\cite{Pang:2018zzo},
\begin{equation}
H(\eta_s) = \exp\left[ -\frac{(|\eta_s| - \eta_0)^2}{2\sigma^2_{\eta_s}}\theta(|\eta_s| - \eta_0) \right].
\end{equation}
The values of $\eta_0$, $\sigma_{\eta_s}$ and the overall normalization $K$ and the initial time $\tau_0$ for different collision systems are listed in Table~\ref{tab:hydro}.

\begin{table}[h]
\centering
\begin{tabular}{|c|c|c|c|c|c|c|}
\hline
	System  & $\sqrt{s_{\rm NN}}$ & $\eta_0$ & $\sigma_{\eta_s}$ & $K$ & $\tau_0$(fm)\\ \hline
	PbPb & 5.02TeV & 1.7 & 2.0  & 154  & 0.6 \\ \hline
	XeXe & 5.44TeV & 1.8 & 2.23 & 155  & 0.6 \\ \hline
	ArAr & 5.85TeV & 1.7 & 2.0  & 160  & 0.6 \\ \hline
	OO & 6.5TeV    & 1.7 & 2.0  & 180  & 0.6 \\ \hline
\end{tabular}
\caption{Parameters for hydrodynamics initial condition for different collision systems.}
\label{tab:hydro}
\end{table}

Using the above setup, we may obtain the initial conditions for hydrodynamics simulation.
In this work, for each centrality we average 5000 Trento events to get the smooth initial entropy distribution as the input for hydrodynamics simulation.
Then we use CLVisc hydrodynamics model~\cite{Pang:2018zzo} to simulate the space-time evolution of the QGP. In the hydrodynamics model, the following equation of motion for the energy-momentum tensor $T^{\mu\nu}$ and the shear stress tensor  $\pi^{\mu\nu}$ are solved in Milne coordinate using the Kurganov-Tadmor (KT) algorithm,
\begin{eqnarray}
& &\partial_{\mu}T^{\mu\nu}  = 0, \\
& &\pi^{\mu\nu} = \eta_v \sigma^{\mu\nu} - \tau_{\pi}\left[\Delta^{\mu\nu}_{\alpha \beta}u^{\lambda}\partial_{\lambda}\pi^{\alpha \beta} + \frac{4}{3}\pi^{\mu\nu}\theta \right],
\end{eqnarray}
where $\sigma^{\mu\nu}$ is the symmetric shear tensor, $\theta$ is the expansion rate.
In this study, the shear-viscosity-to-entropy-density-ratio is chosen to be $\eta_v/s = 0.16$ and the relaxation coefficient is $\tau_{\pi} = \frac{3\eta_v}{sT_m}$, and the equation of state is taken from the {\rm s95-pce} parameterization.
After hydrodynamic evolution, the QGP medium will be converted to hadrons according to the Cooper-Frye formula where the switching temperature is taken as $T_{\rm sw}=137$ MeV.
Note that the soft hadrons produced from the freezeout hypersurface of the fluid have negligible contribution to the nuclear modification of hard hadrons with $p_T > 6-8$~GeV explored in this paper~\cite{Zhao:2021vmu}.

\begin{figure*}[tbh]
\includegraphics[width=0.450\textwidth]{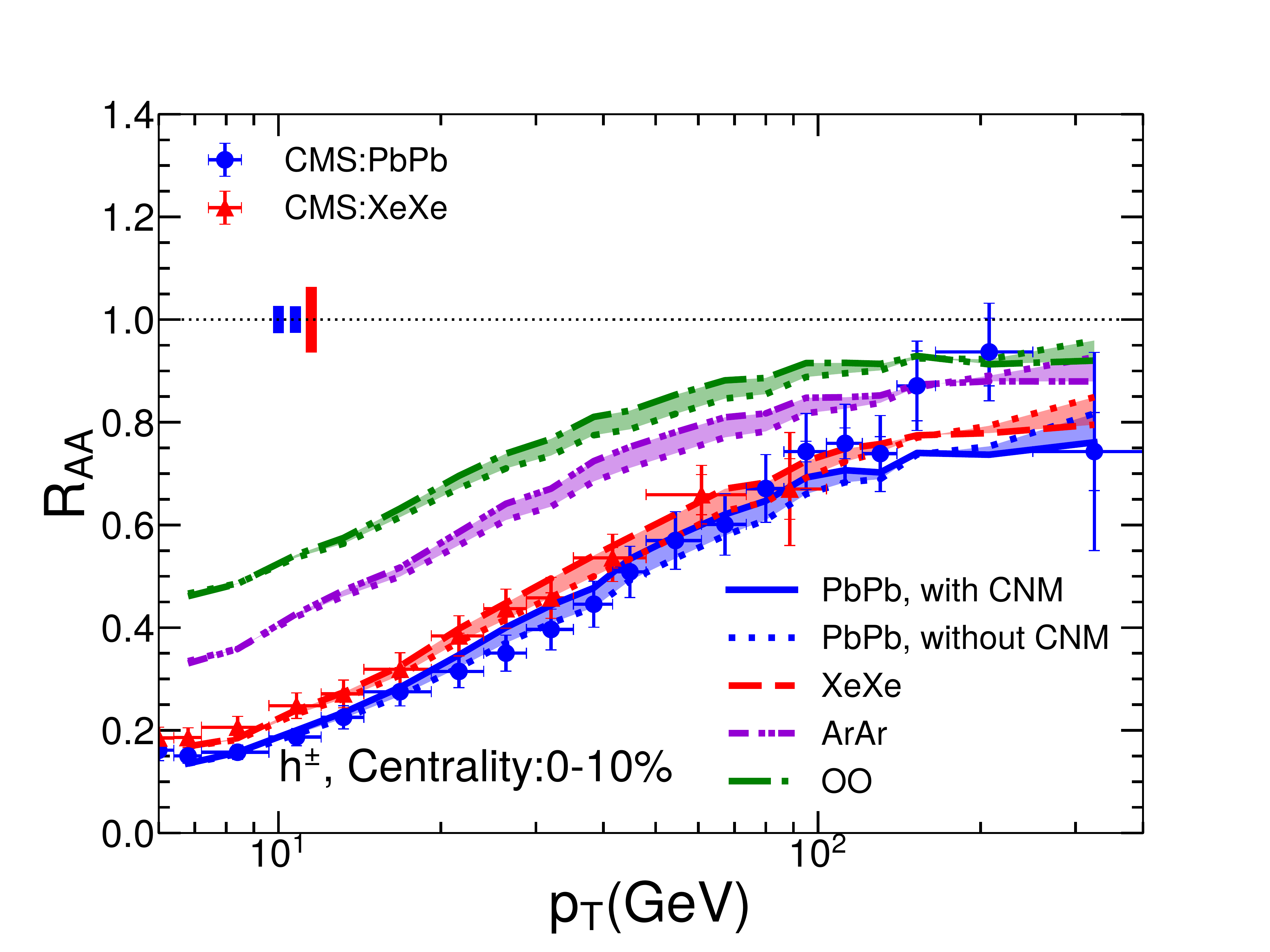}
\includegraphics[width=0.450\textwidth]{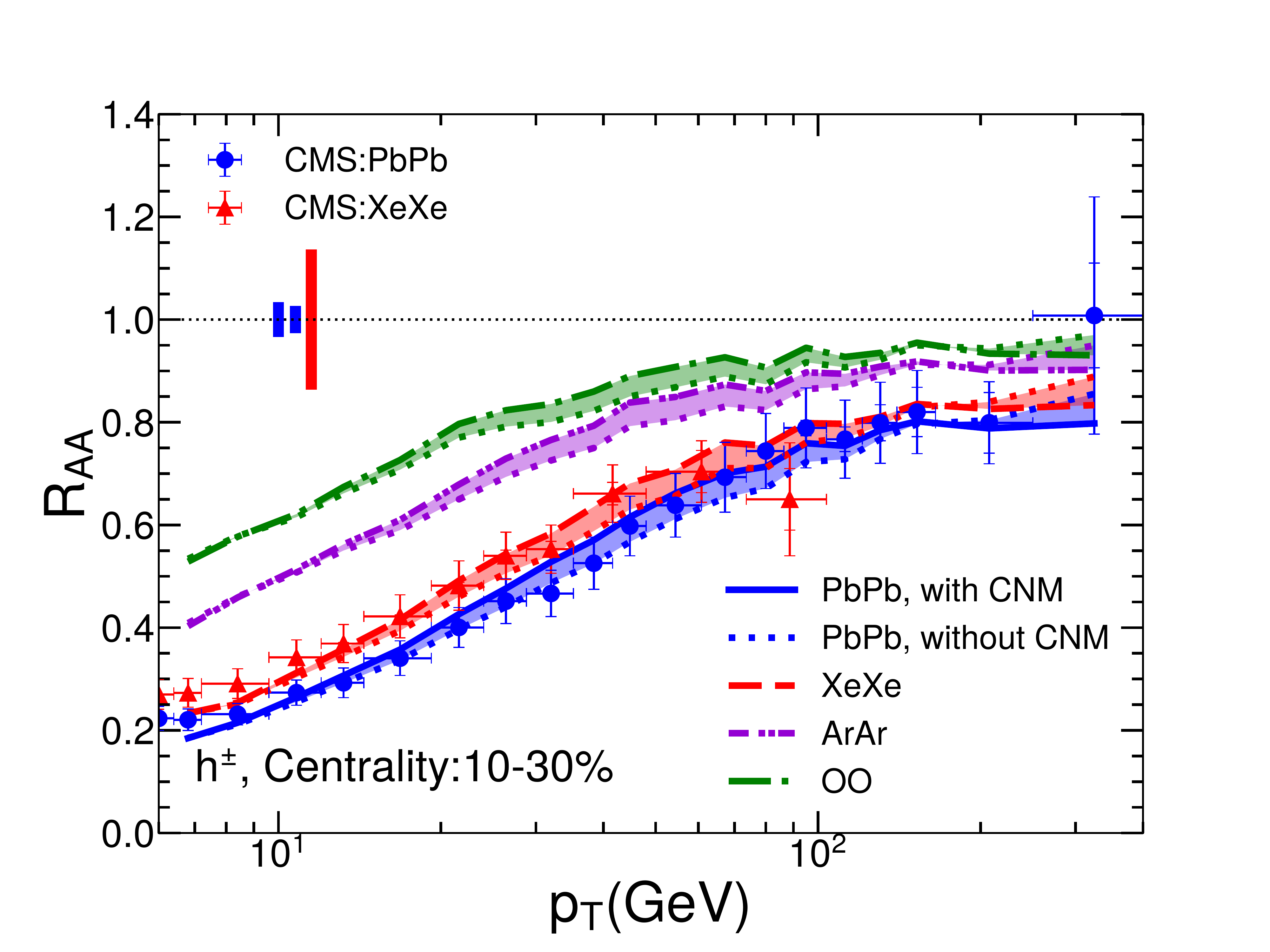}
\includegraphics[width=0.450\textwidth]{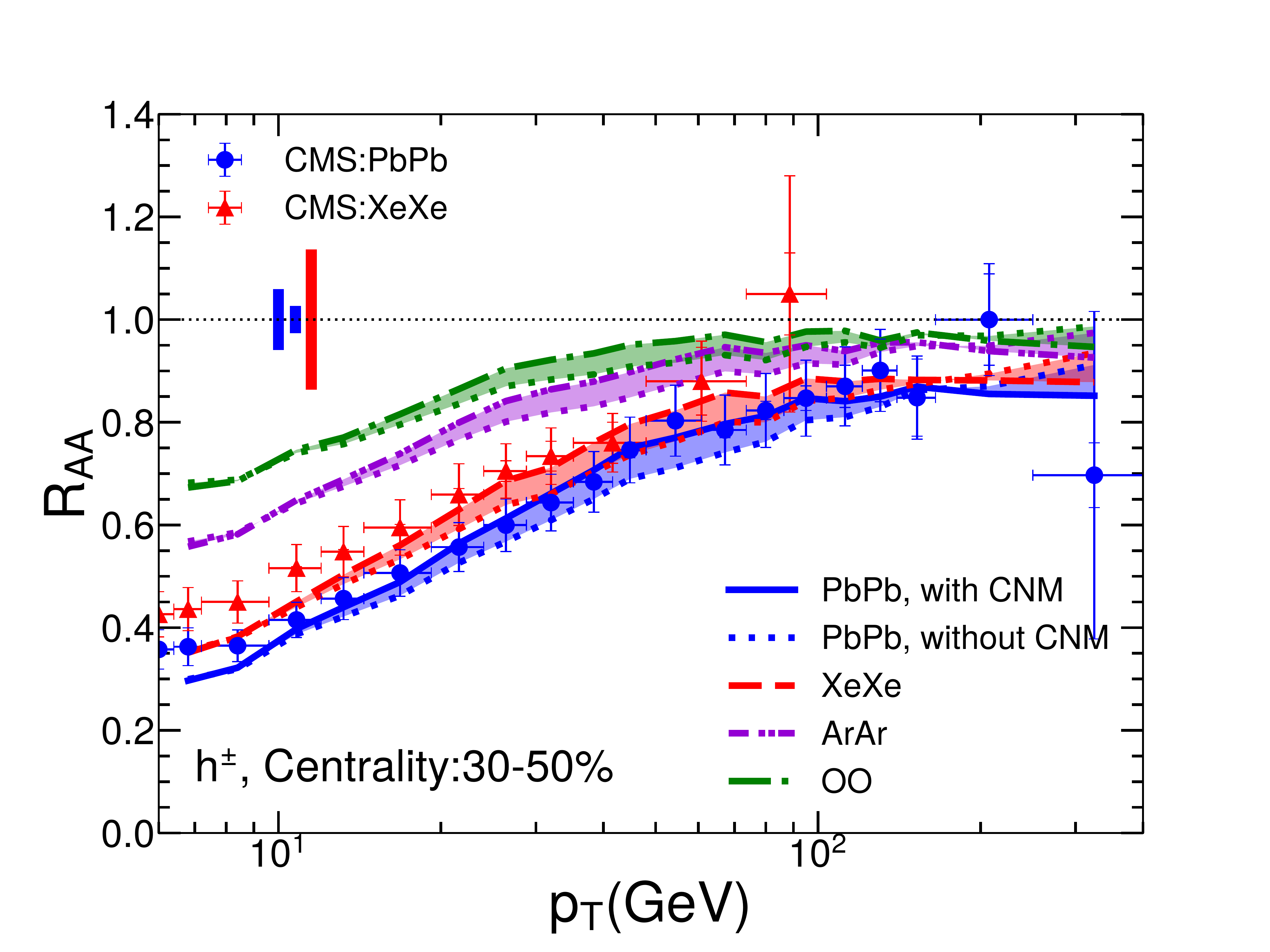}
\includegraphics[width=0.450\textwidth]{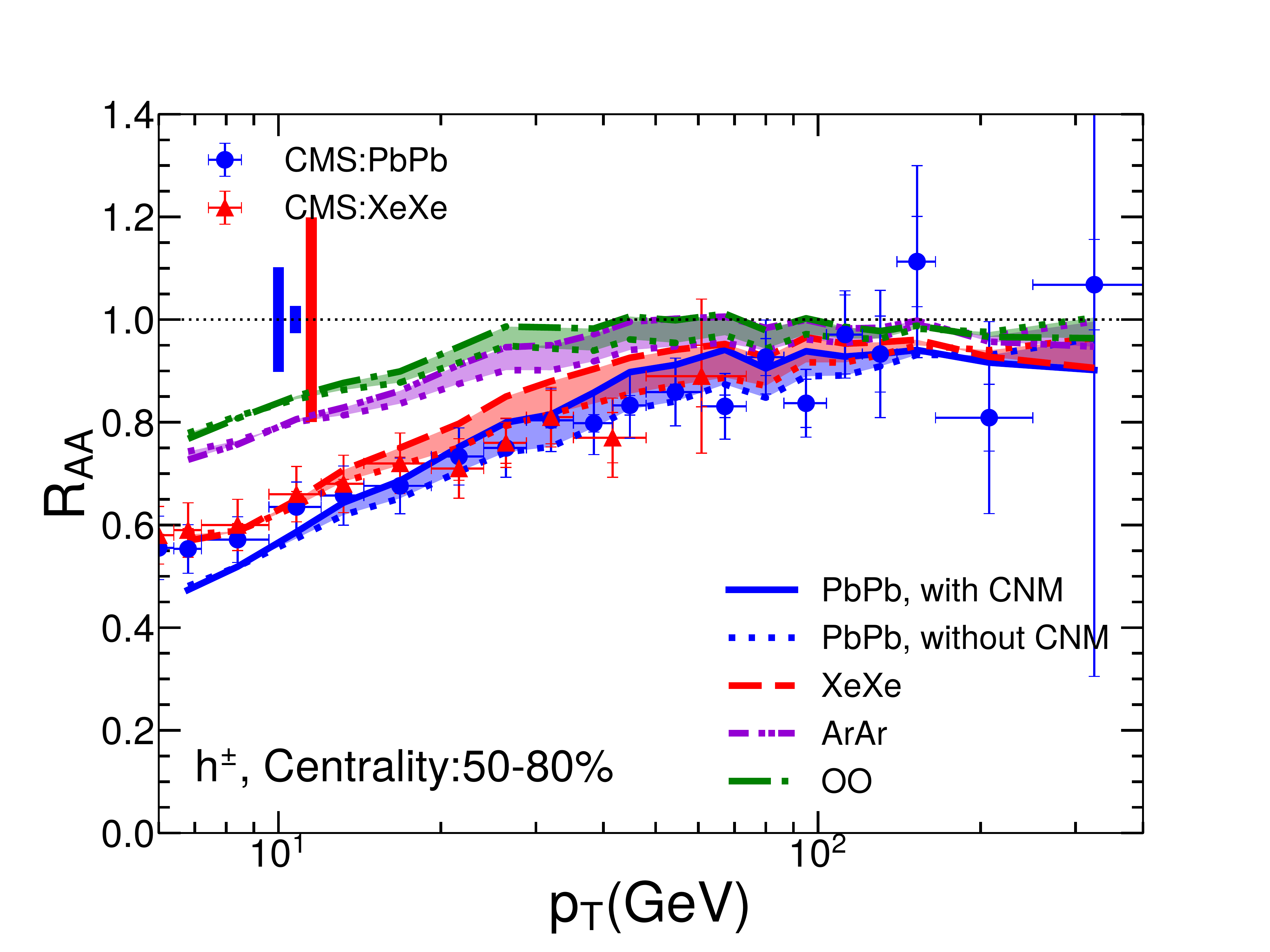}
	\caption{$R_{AA}$ of charged hadrons as a function of $p_T$ in four collision systems and in four centrality classes. The data are taken from CMS Collaboration \cite{Khachatryan:2016odn, Sirunyan:2018eqi}.}
	\label{RAA_pT_h}
\includegraphics[width=0.450\textwidth]{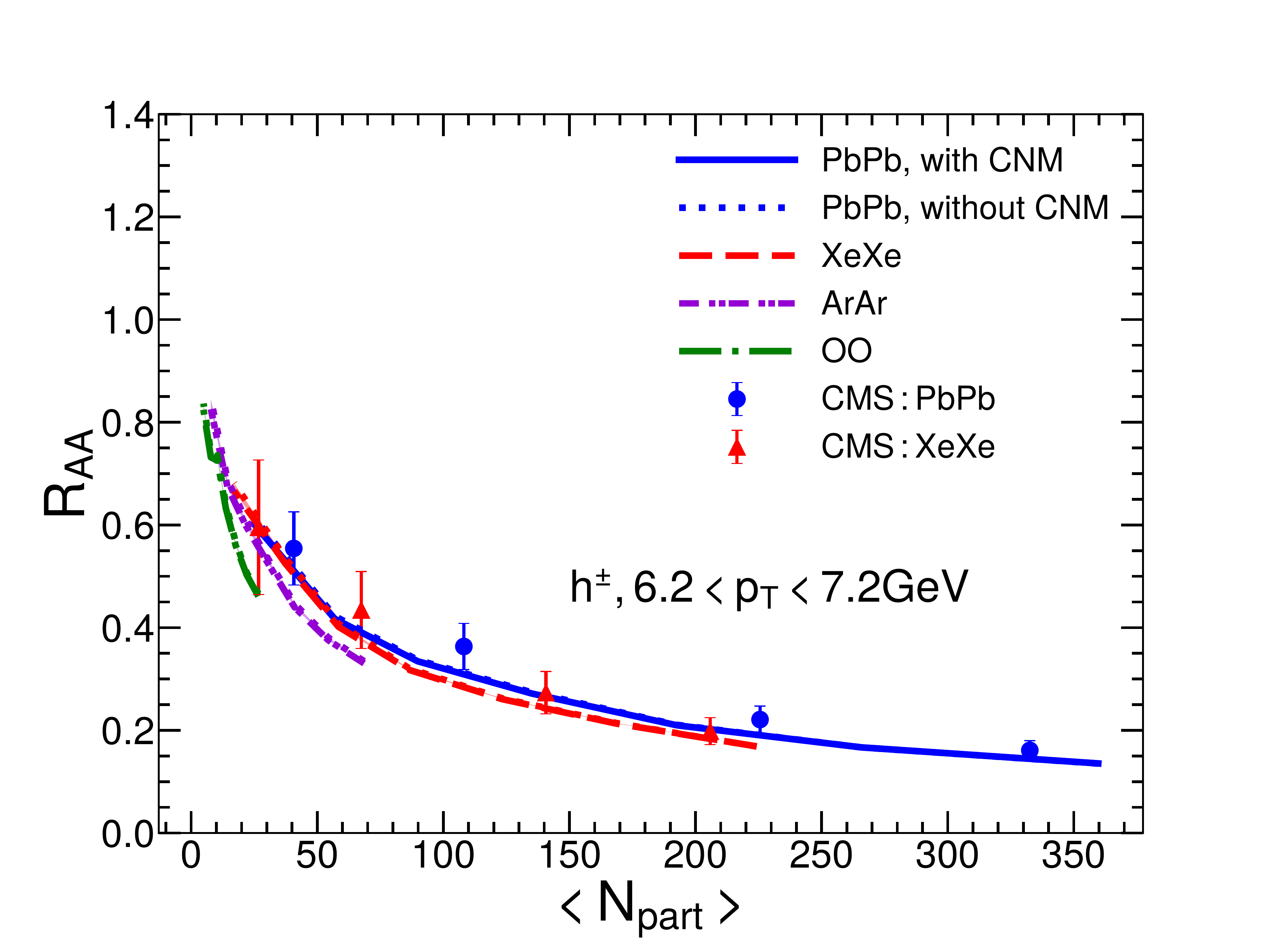}
\includegraphics[width=0.450\textwidth]{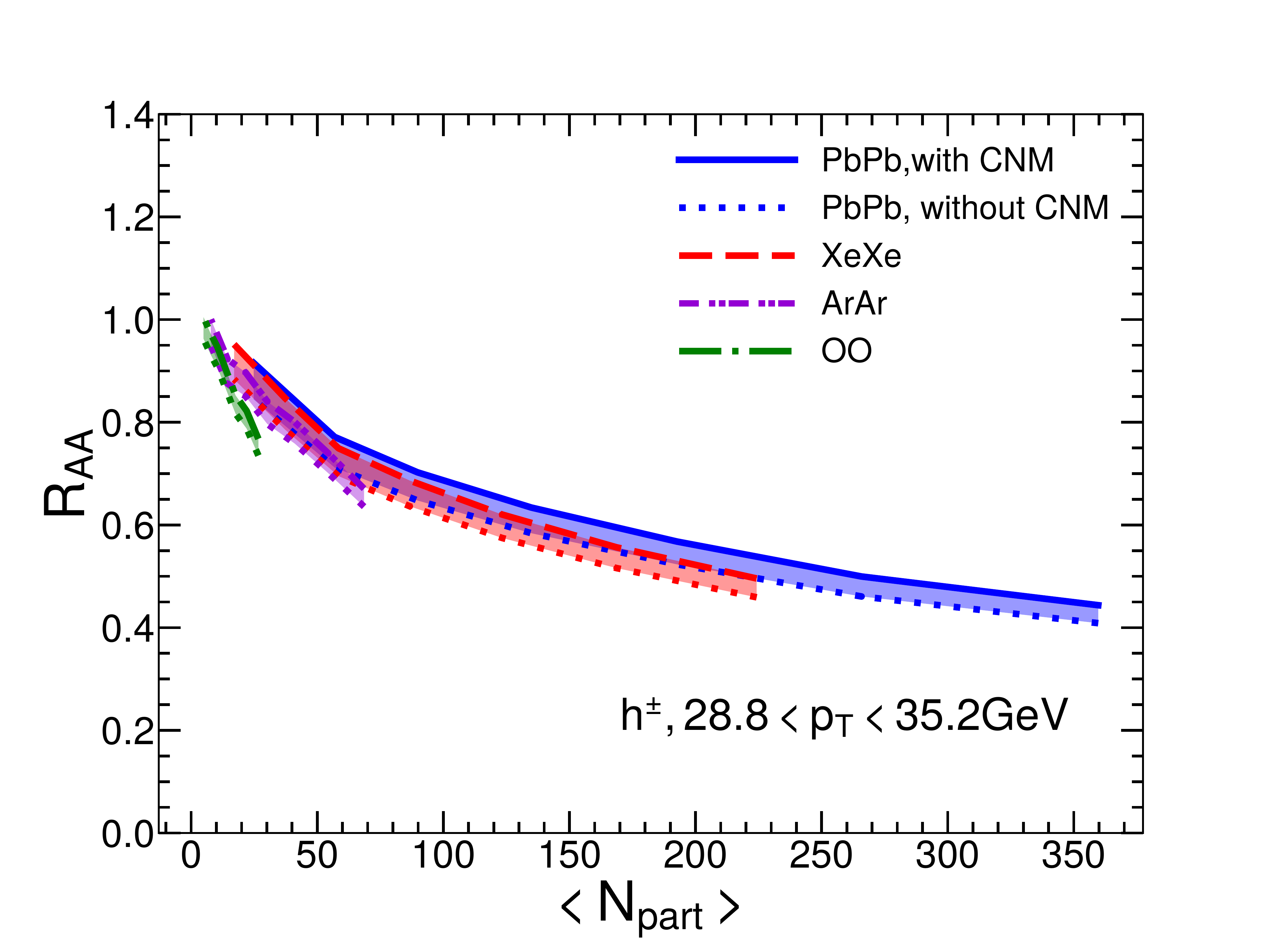}
	\caption{$R_{AA}$ of charged hadrons as a function of $\langle N_{\rm part}\rangle$ in four collision systems and for two $p_T$ bins. The data are taken from CMS Collaboration \cite{Sirunyan:2018eqi}.}
	\label{RAA_Npart_h}
\end{figure*}

\begin{figure*}[tbh]
\includegraphics[width=0.450\textwidth]{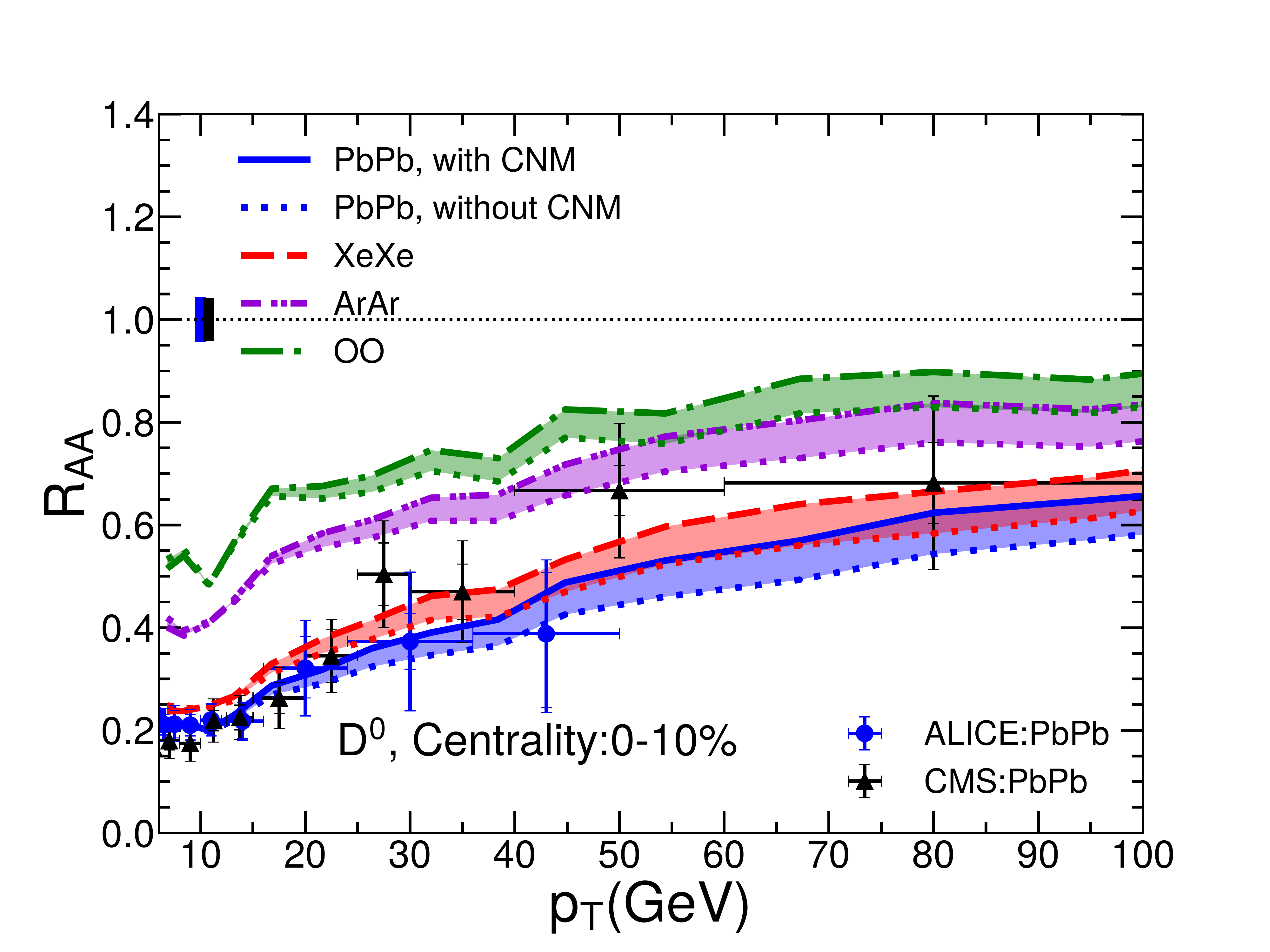}
\includegraphics[width=0.450\textwidth]{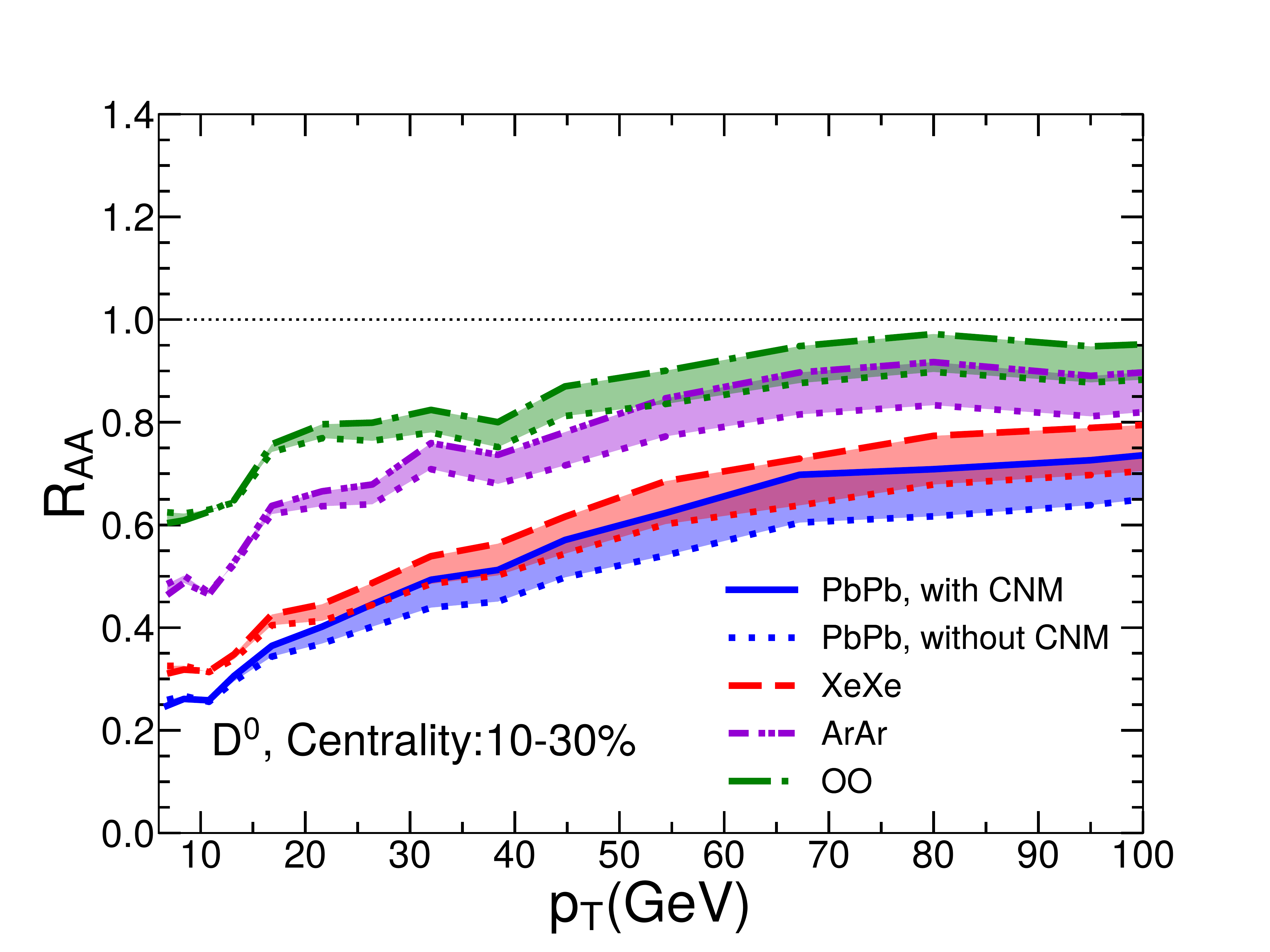}
\includegraphics[width=0.450\textwidth]{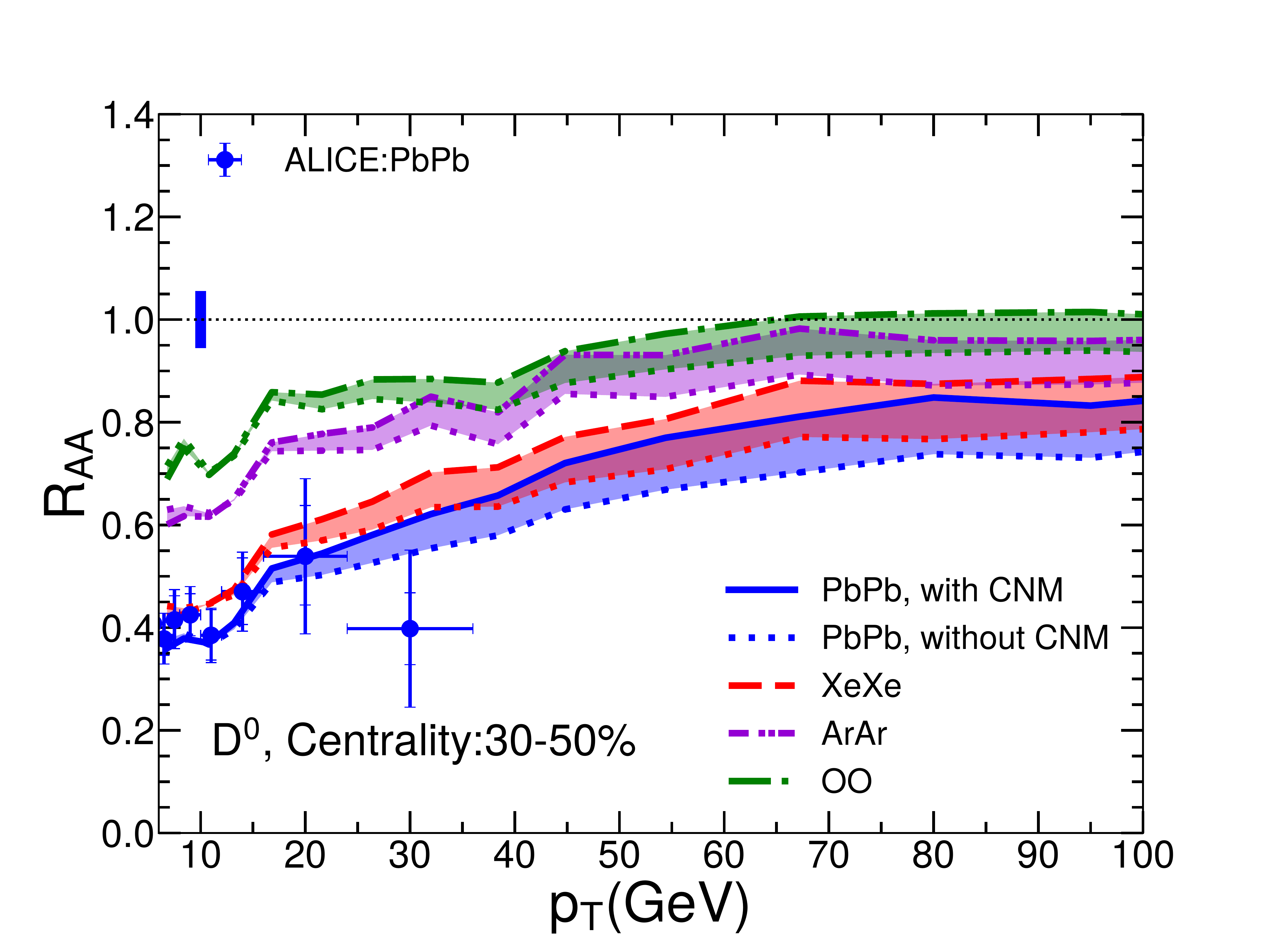}
\includegraphics[width=0.450\textwidth]{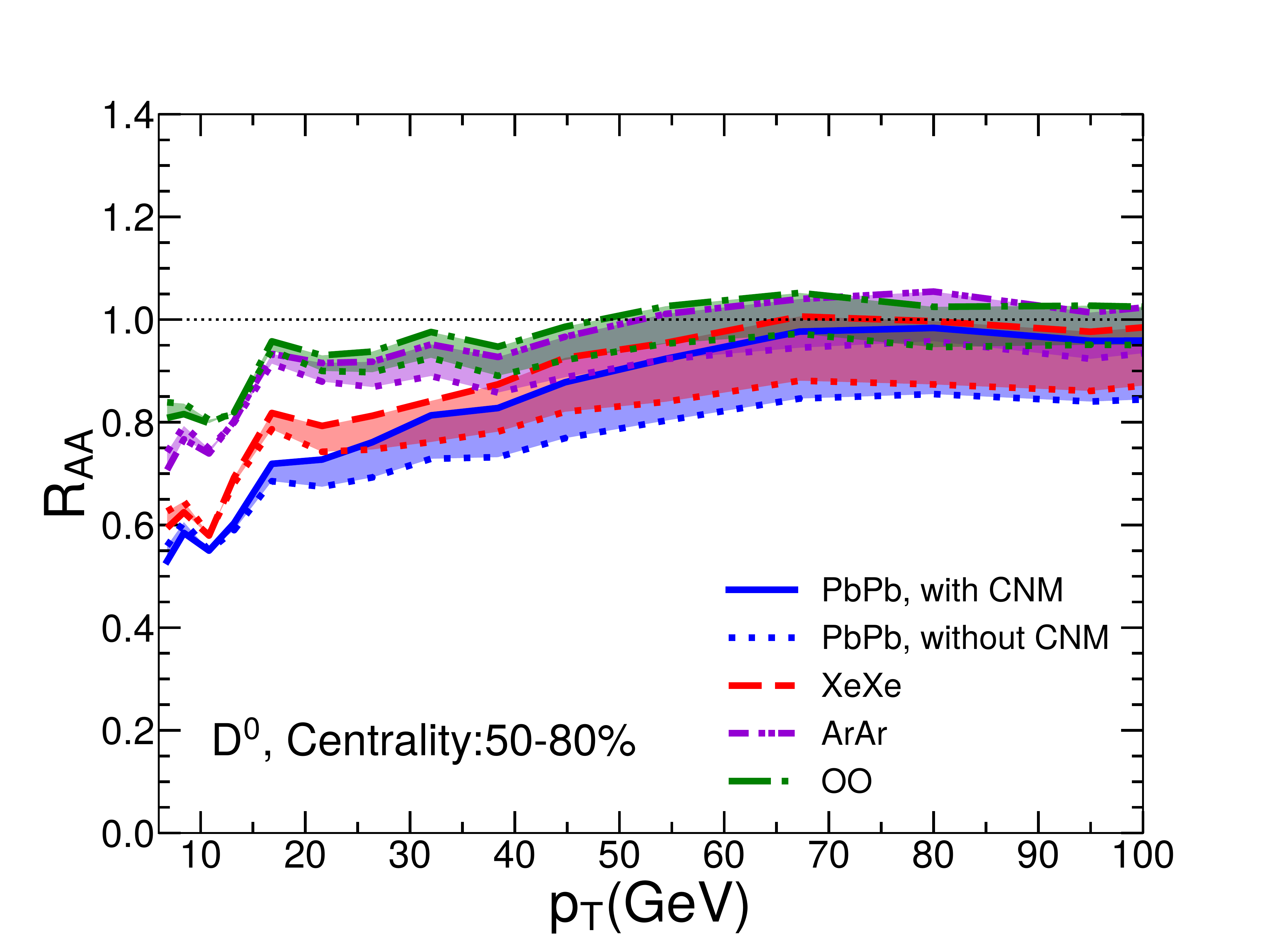}
	\caption{$R_{AA}$ of prompt $D$ mesons as a function of $p_T$ in four collision systems and in four centrality classes. The data are taken from CMS and ALICE Collaborations \cite{Sirunyan:2017xss, Acharya:2018hre}.}
	\label{RAA_pT_D}
\includegraphics[width=0.450\textwidth]{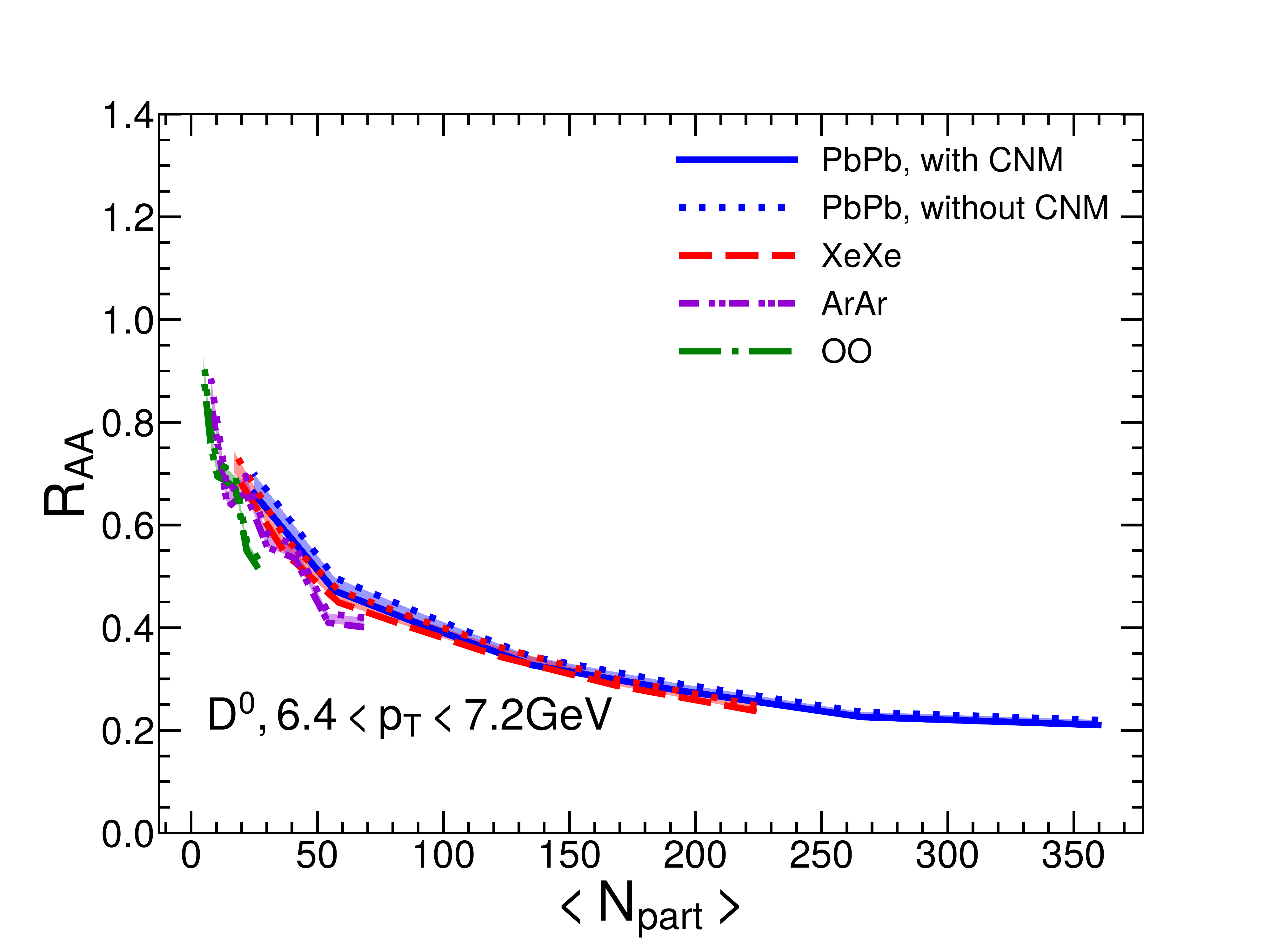}
\includegraphics[width=0.450\textwidth]{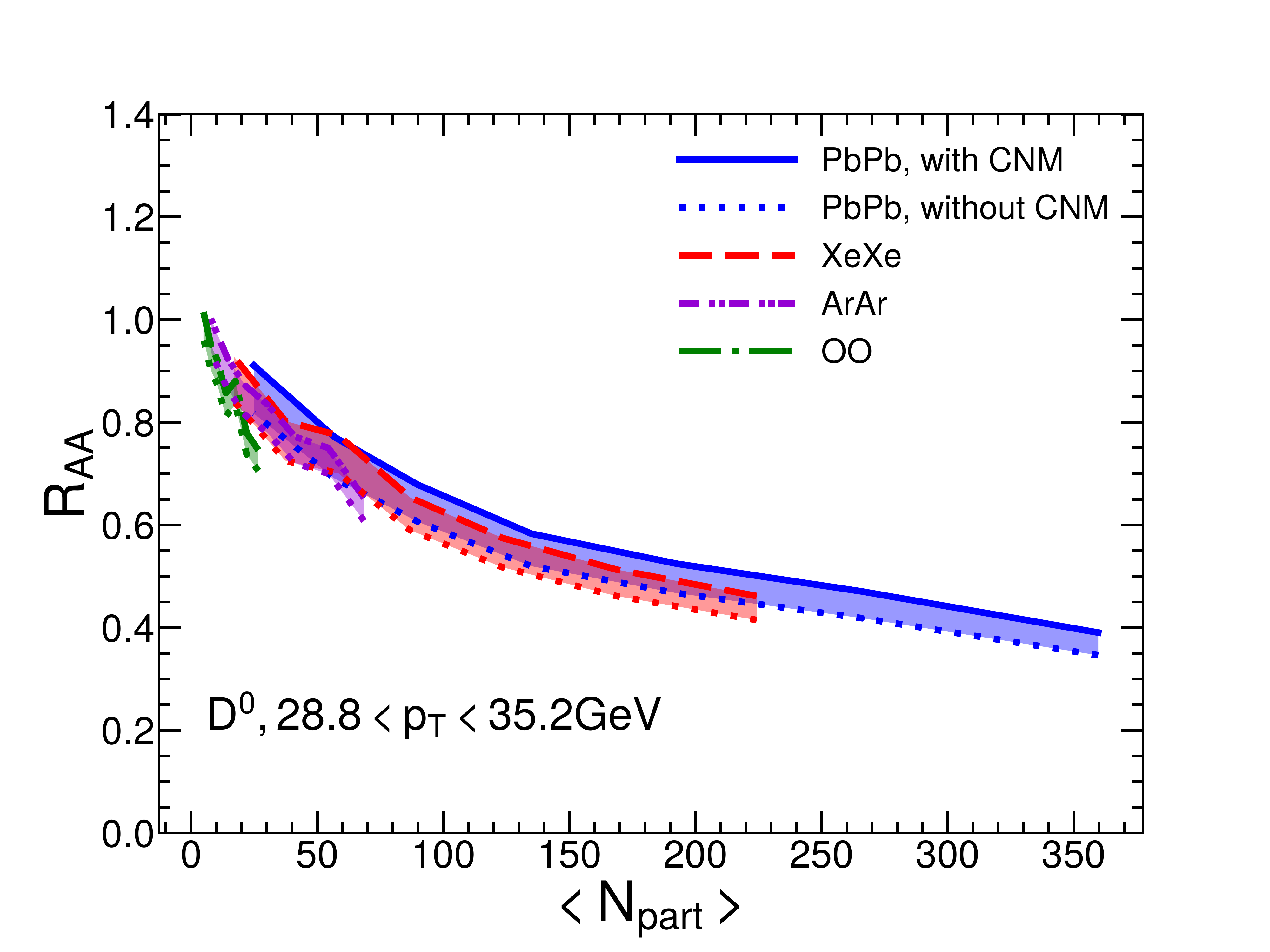}
\caption{$R_{AA}$ of prompt $D$ mesons as a function of $\langle N_{\rm part}\rangle$ in four collision systems for two $p_T$ bins. }
	\label{RAA_Npart_D}
\end{figure*}

\begin{figure*}[tbh]
\includegraphics[width=0.450\textwidth]{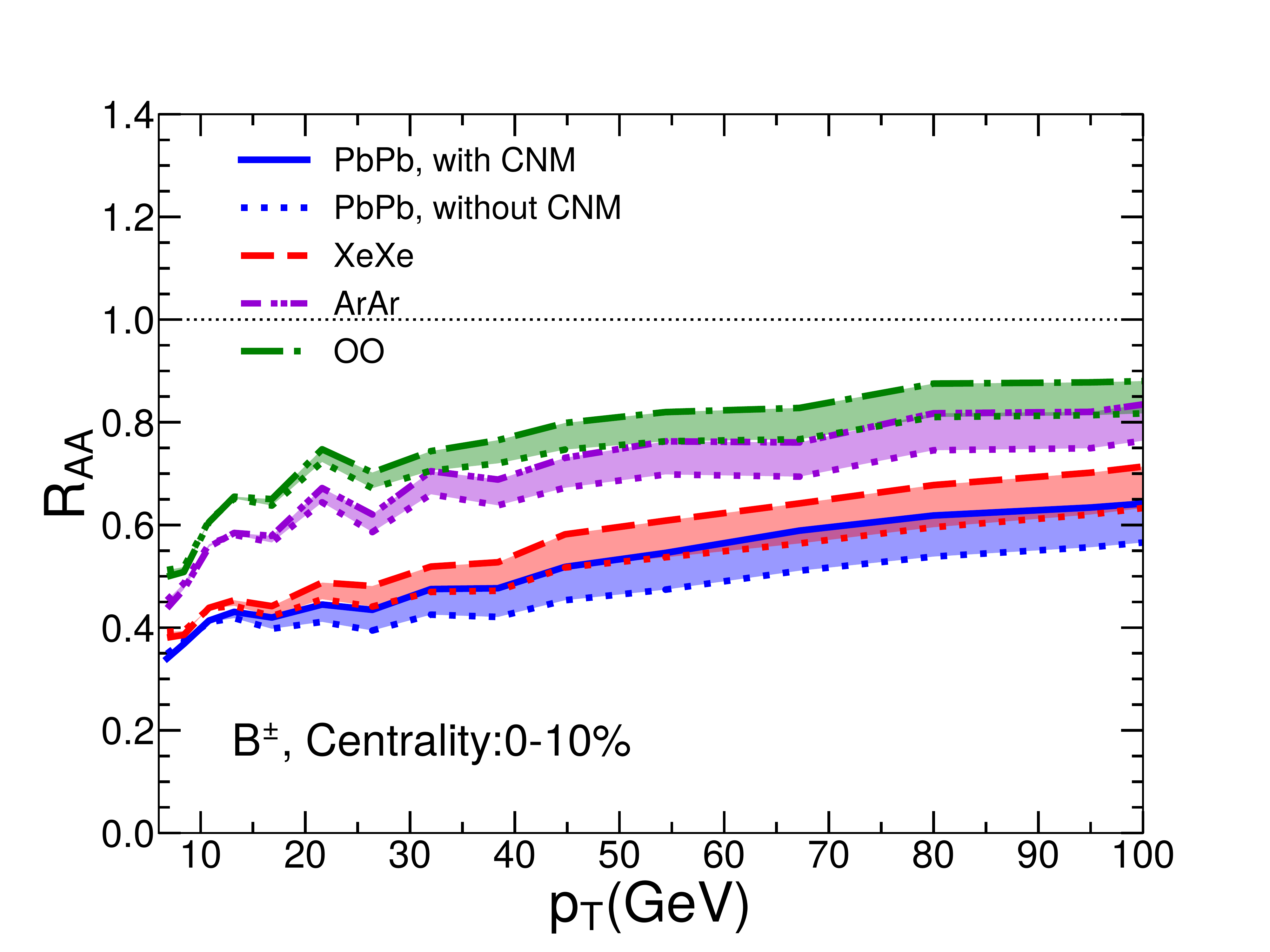}
\includegraphics[width=0.450\textwidth]{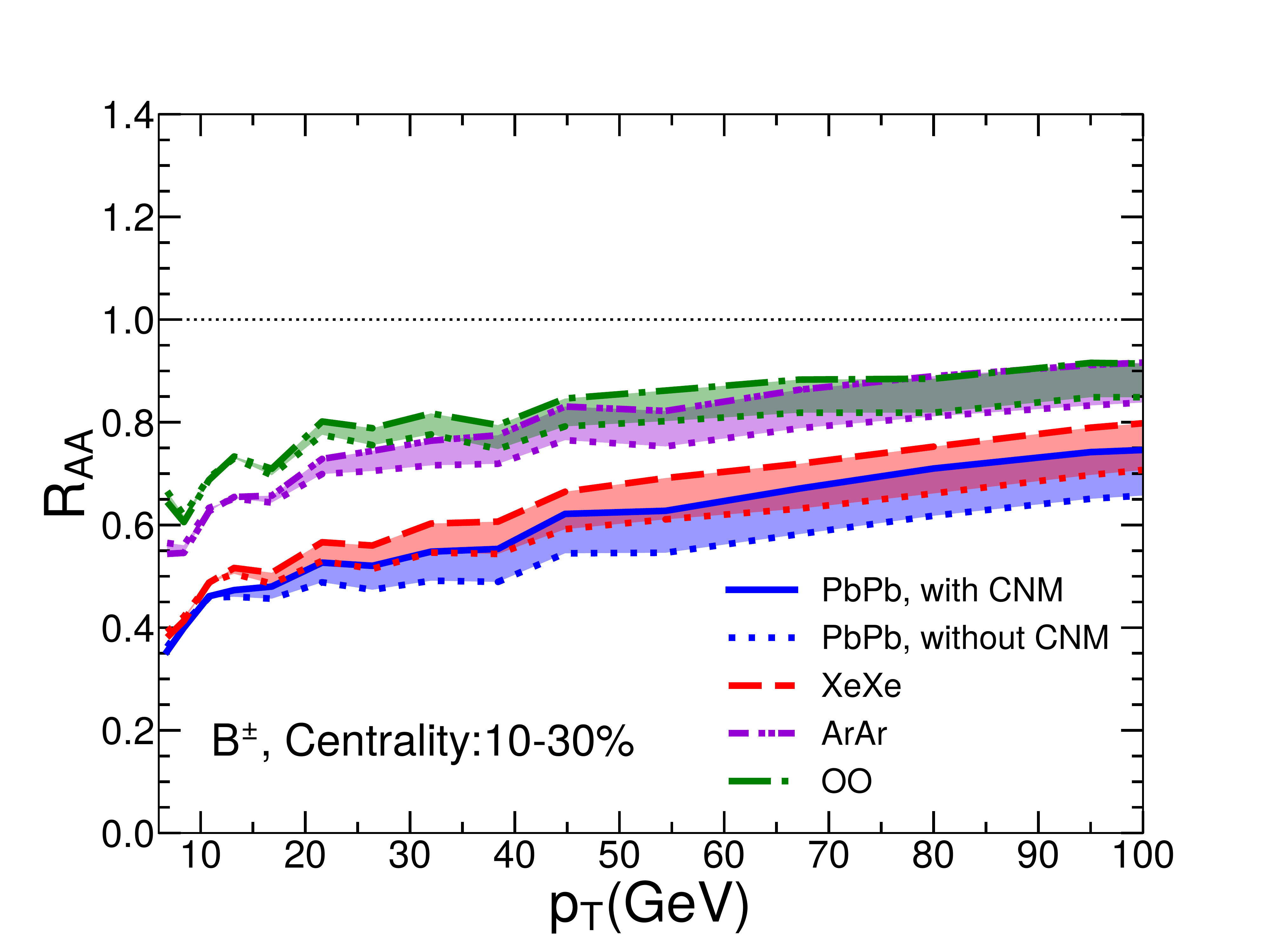}
\includegraphics[width=0.450\textwidth]{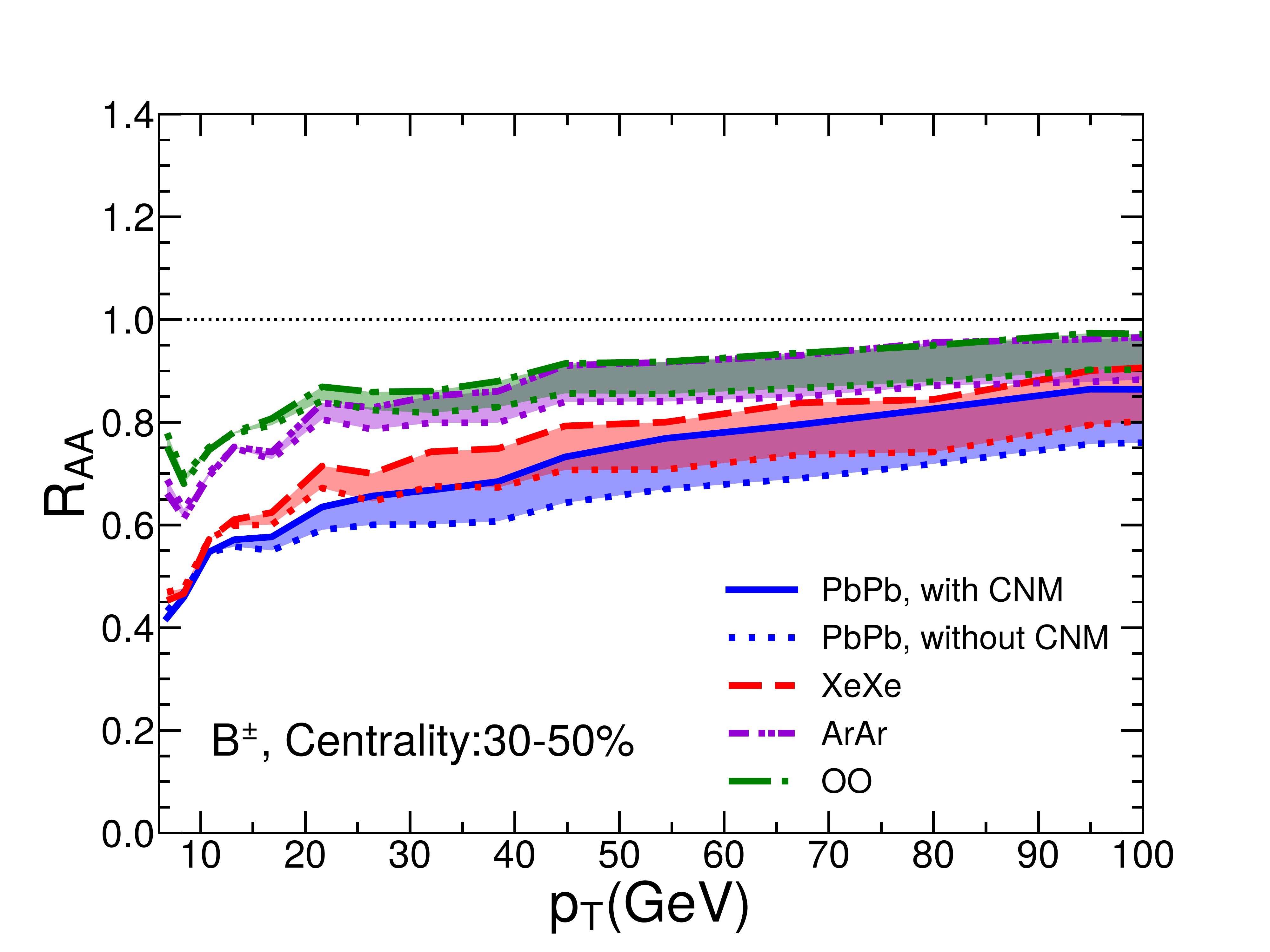}
\includegraphics[width=0.450\textwidth]{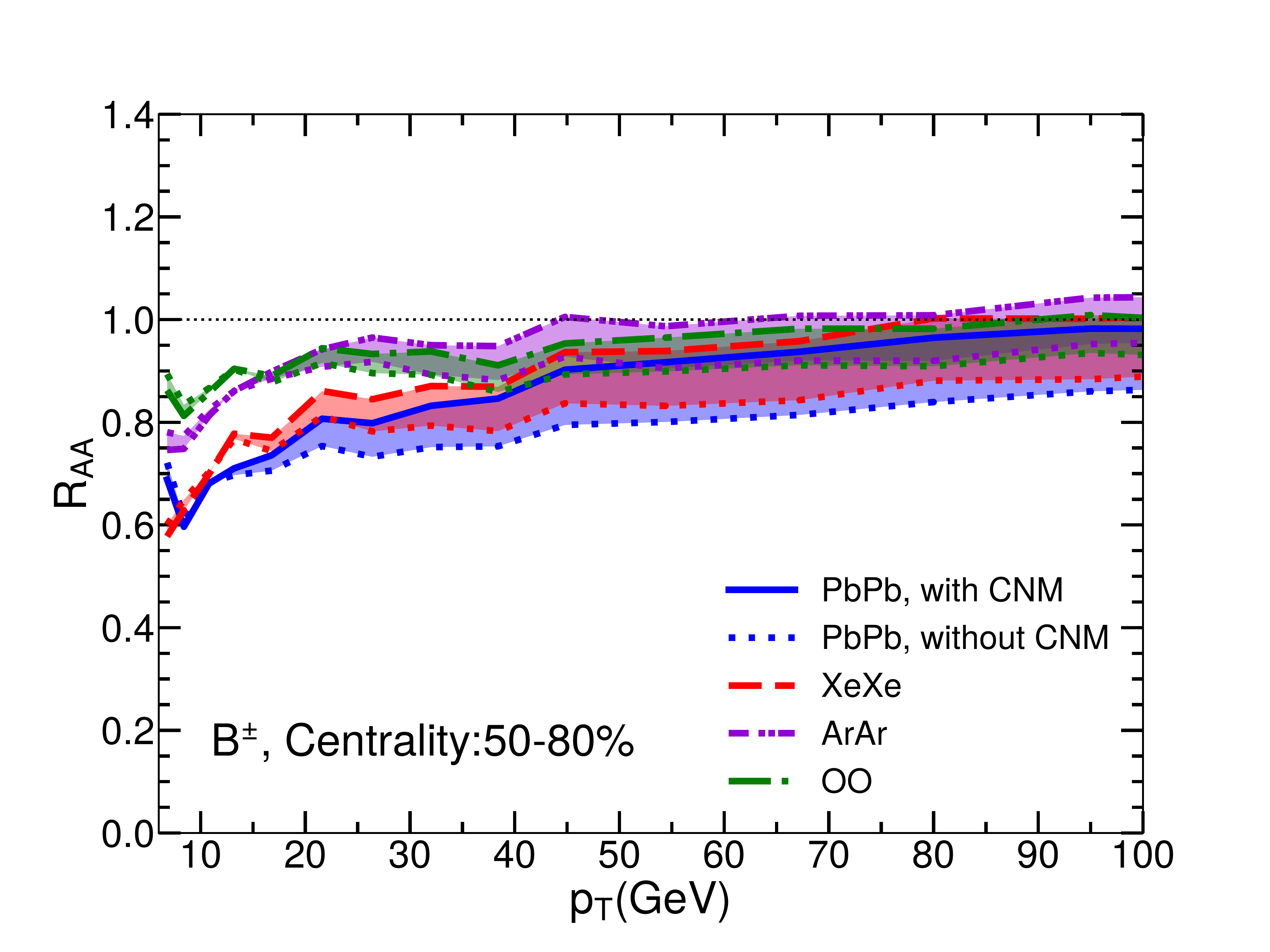}
\caption{$R_{AA}$ of $B$ mesons as a function of $p_T$ in four collision systems and in four centrality classes. }
	\label{RAA_pT_B}
\includegraphics[width=0.450\textwidth]{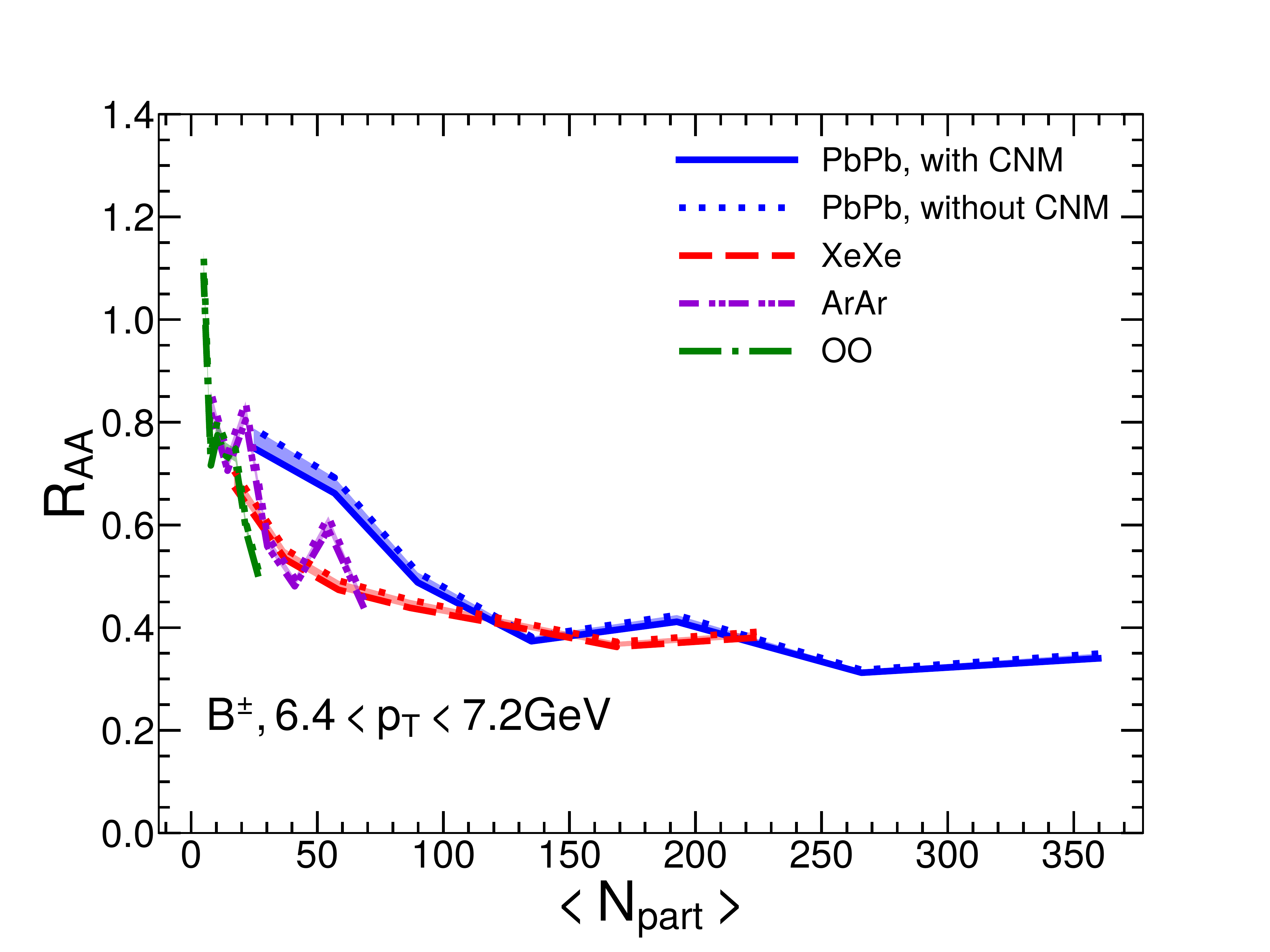}
\includegraphics[width=0.450\textwidth]{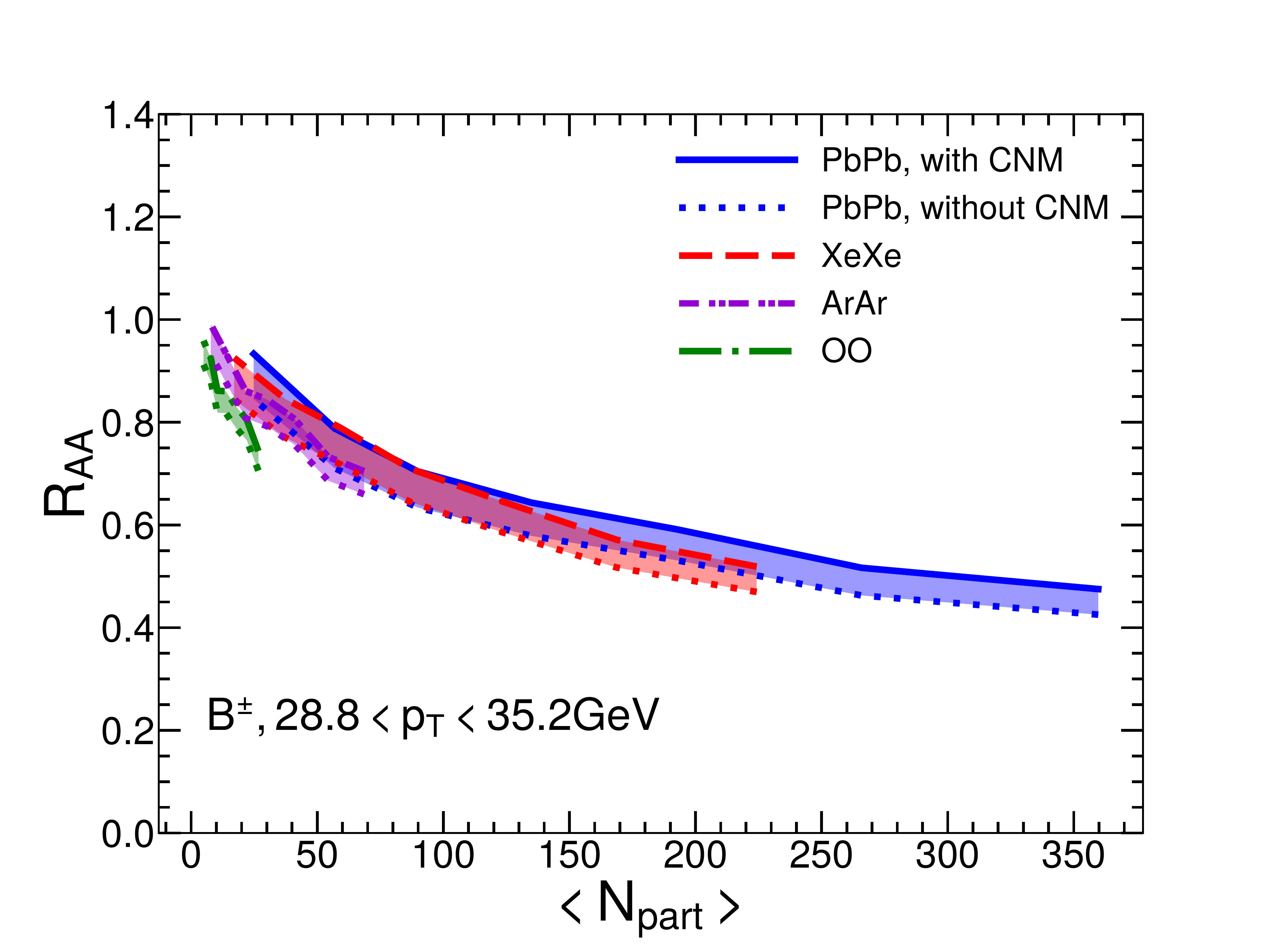}
\caption{$R_{AA}$ of $B$ mesons as a function of $\langle N_{\rm part}\rangle$ in four collision systems for two $p_T$ bins. }
	\label{RAA_Npart_B}
\end{figure*}

\begin{figure*}[tbh]
\includegraphics[width=0.450\textwidth]{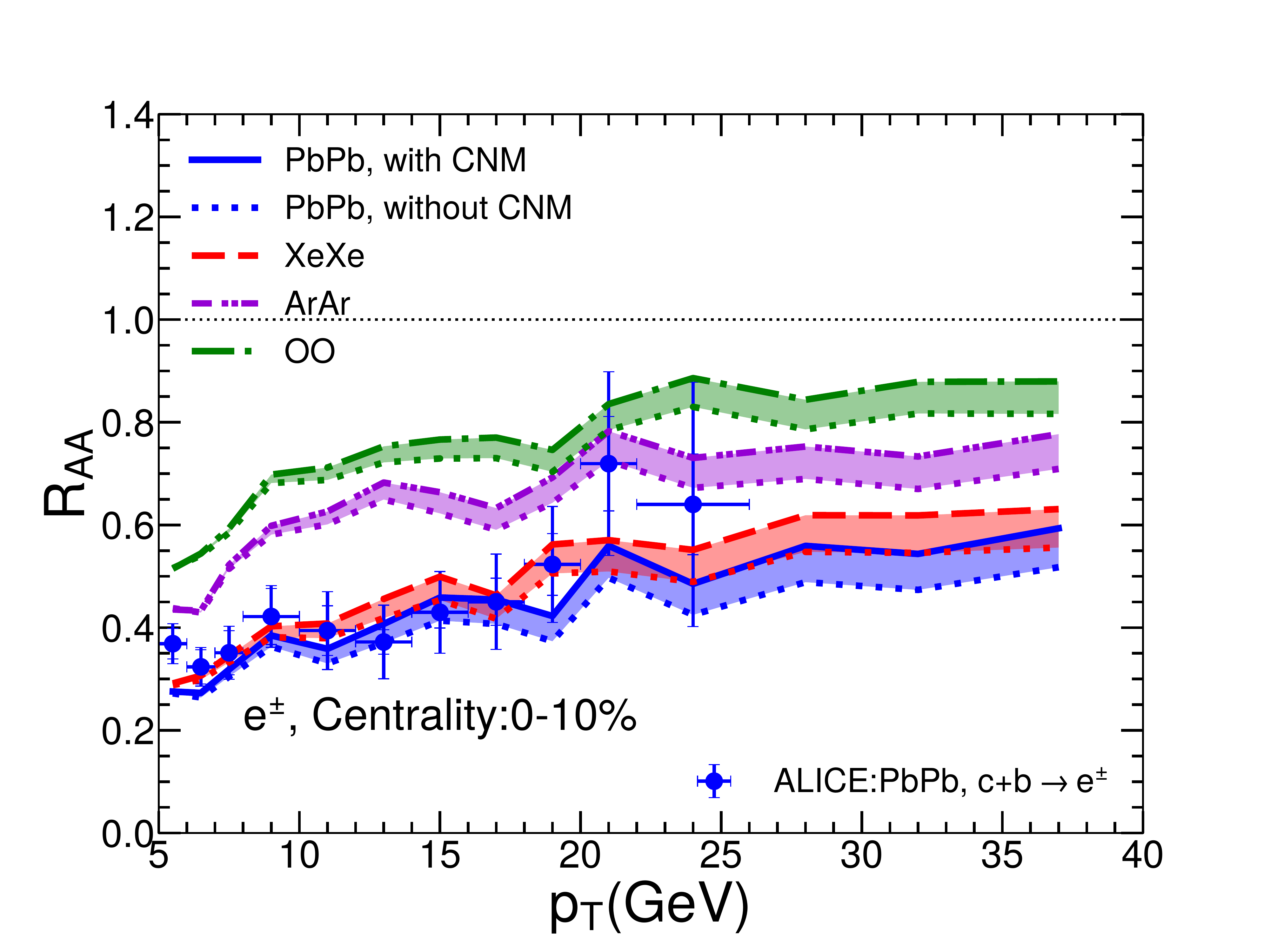}
\includegraphics[width=0.450\textwidth]{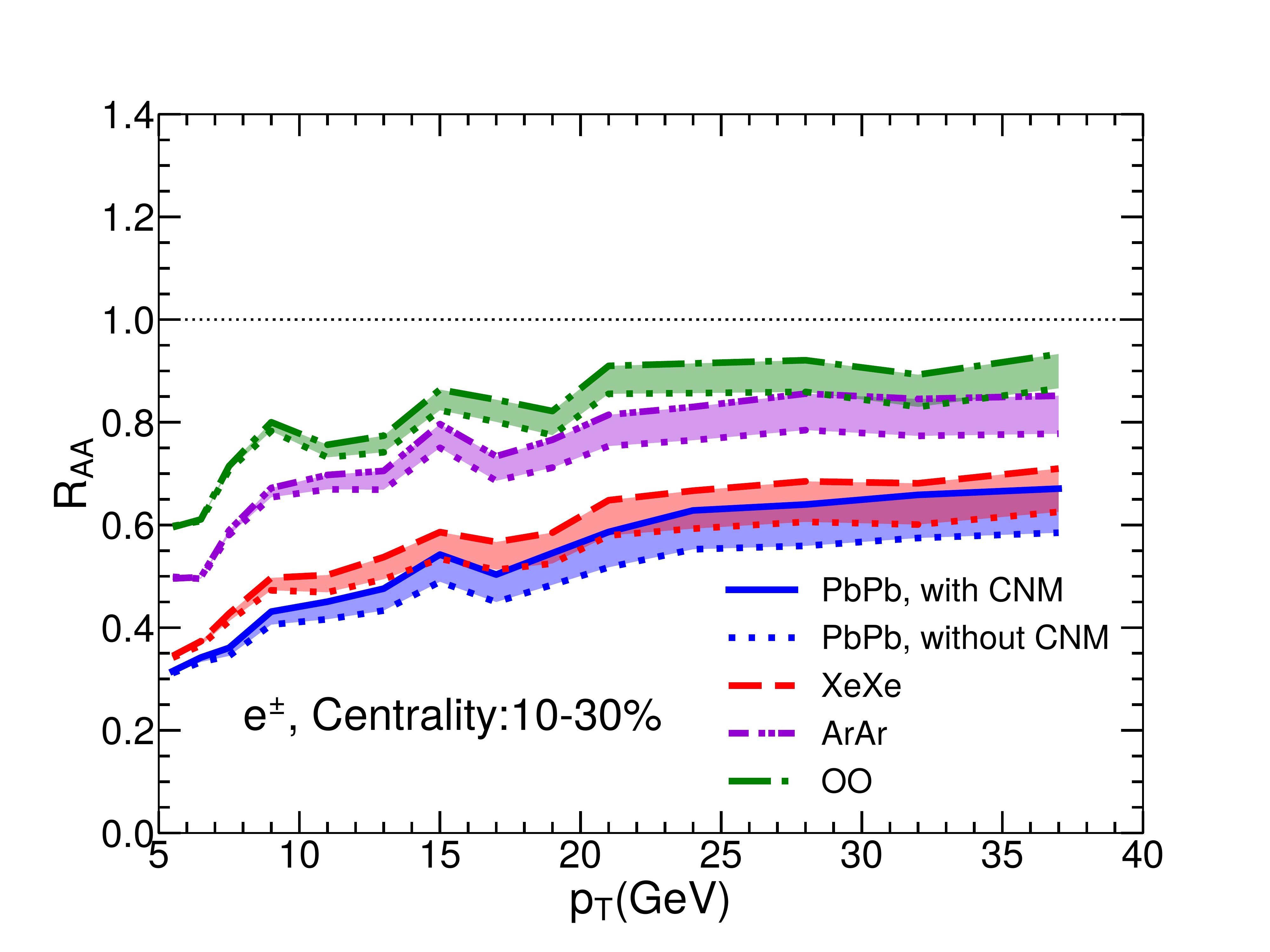}
\includegraphics[width=0.450\textwidth]{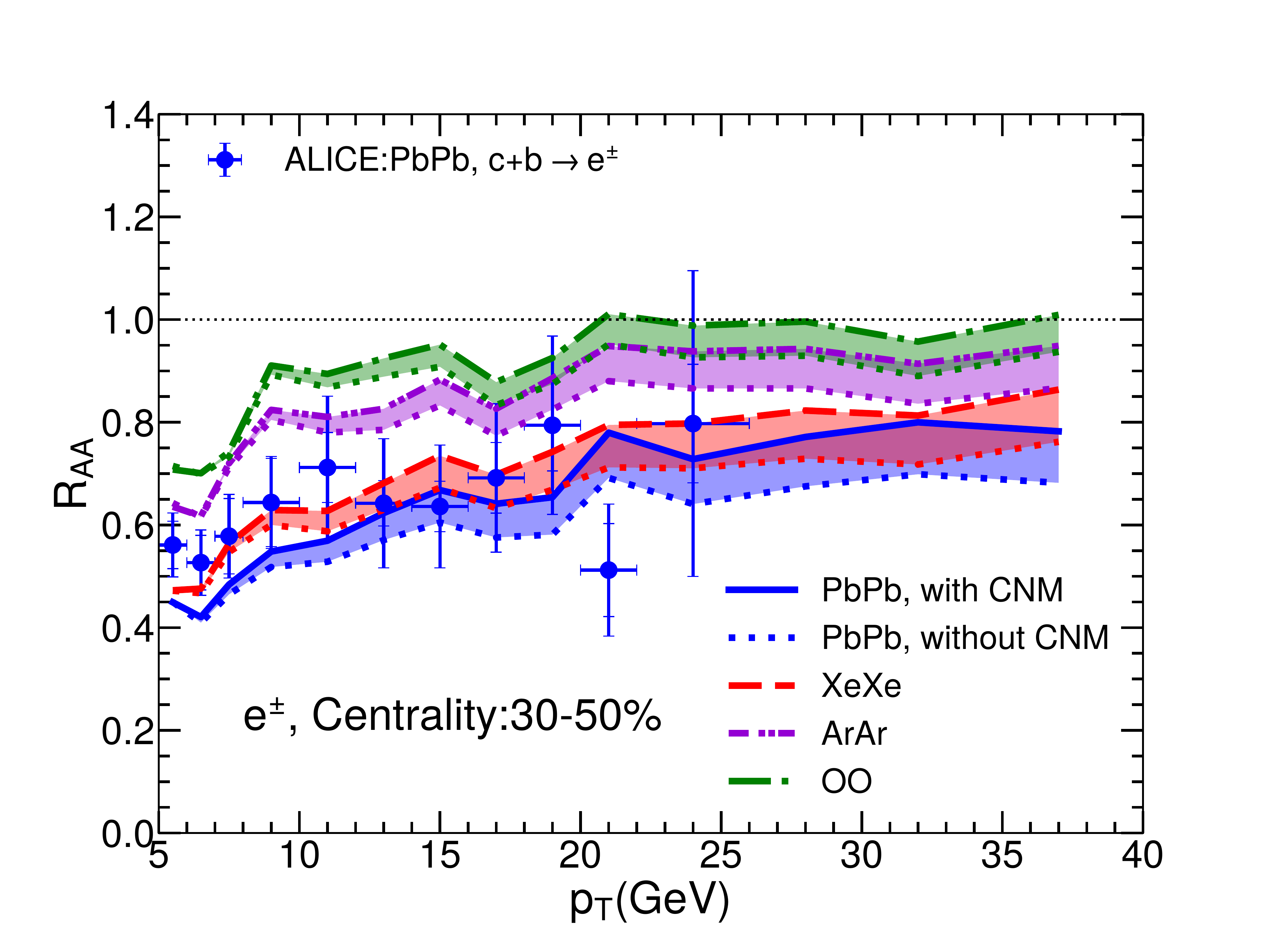}
\includegraphics[width=0.450\textwidth]{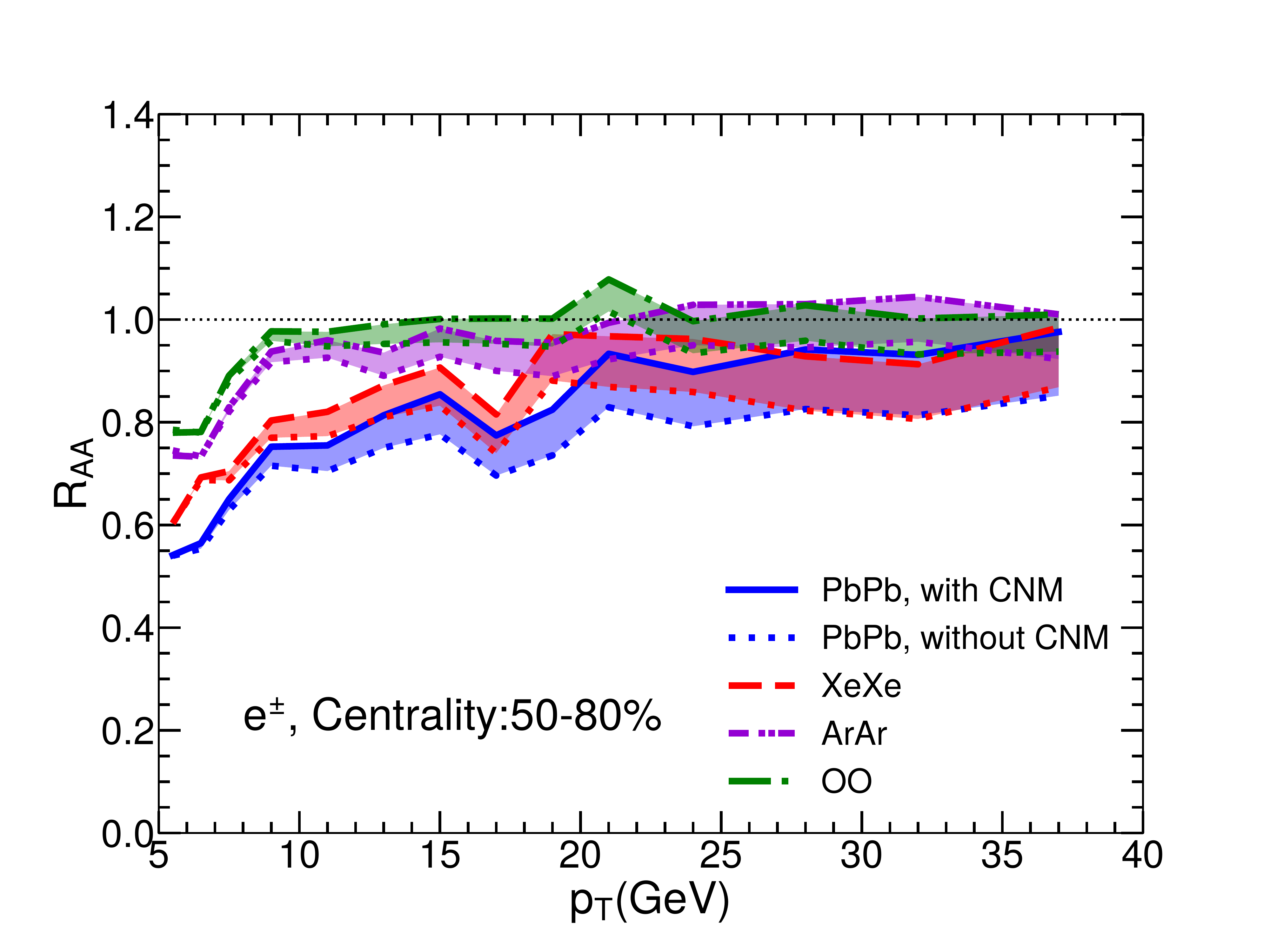}
	\caption{$R_{AA}$ of heavy flavor decayed electrons ($D,B \to e$) as a function of $p_T$ in four collision systems and in four centrality classes. The data are taken from ALICE Collaboration \cite{Acharya:2019mom}. }
\label{RAA_pT_e}
\end{figure*}


\section{Numerical results}

In Fig. \ref{RAA_pT_h}, we show the nuclear modification factor $R_{AA}$ as a function of transverse momentum $p_T$ for charged hadrons in Pb+Pb collisions at 5.02A~TeV, Xe+Xe collisions at 5.44A~TeV, Ar+Ar collisions at 5.85A~TeV and O+O collisions at 6.5A~TeV at the LHC.
Four panels show four different centralities: 0-10\%, 10-30\%, 30-50\% and 50-80\%.
The bands show the uncertainties due to the cold nuclear matter effect: solid/dashed/dash-dotted/dash-dot-dotted curves represent our calculation results with CNM while dotted curves are corresponding results without CNM.
Note that as there is no nuclear PDF for $^{40}$Ar from EPS09 parameterizations, we take the nuclear PDF of $^{40}$Ca as an approximation in the calculation.
We have checked that  for Pb+Pb collisions, the use of the impact parameter dependent nPDF -- EPS09s~\cite{Helenius:2012wd} makes the CNM uncertainty band of $R_{AA}$ narrower in peripheral collisions, but has small effect in central collisions. One possible reason is that central collisions have much larger binary collision number than peripheral collisions, thus have much larger contribution when averaging over the centrality (impact parameter) for the CNM effect.

From the plots, one can see that our jet quenching model can provide a very good description of the transverse momentum and centrality dependences of $R_{AA}$ for charged hadrons in  Pb+Pb and Xe+Xe collisions as measured by CMS Collaboration.
Using the same setups, we predict charged hadron $R_{AA}$ for smaller collision systems, such as Ar+Ar and O+O collisions, as a function of $p_T$ and centrality.
One can clearly see the system size dependence for $R_{AA}$, i.e., $R_{AA}$ is larger for smaller systems, due to smaller jet energy loss effect.
The system size dependence can also been seen from the centrality dependence: $R_{AA}$ becomes closer to unity when moving from central to peripheral collisions.

It is very interesting to see that the jet quenching effect on $R_{AA}$ is still very significant in central O+O collisions: $R_{AA}$ at $p_T \sim$~8-10~GeV can reach about 0.5.
Such quenching effect should be large enough to be observed in the future O+O collisions at the LHC.
The confirmation of our result in the future can serve as the bridge to search for the QGP signal in even smaller collision systems like proton-nucleus collisions.
It is worth noting that the experimental bias in event selection may lead to sizable deviation from unity for $R_{AA}$ in peripheral collisions~\cite{Acharya:2018njl, Loizides:2017sqq}.

In order to see the system size dependence more clearly, we plot in Fig. \ref{RAA_Npart_h} the nuclear modification factor $R_{AA}$ as a function of $N_{\rm part}$ for four different collision systems (as shown by four different curves).
The left figure shows the result for $p_T = (6.7, 7.7)$~GeV and the right figure for $p_T = (26, 30)$~GeV.
Our model can describe the experimental data for Pb+Pb collisions and Xe+Xe collisions quite well.
One now sees that $R_{AA}$ is a decreasing function of $N_{\rm part}$, which clearly demonstrates the system size effect on the jet energy loss and nuclear modification factor $R_{AA}$.
As one goes from larger to smaller collision systems, $R_{AA}$ increases and tends to reach unity.
One also observes that for the same values of $N_{\rm part}$, there exist some differences in $R_{AA}$ from four different collision systems due to different collision energies.
We have checked that the effect of the different initial parton spectra is negligible (as the collision energies are not very different).
This means that different initial geometries and hydrodynamic evolution profiles play more important roles for such difference.

In Fig. \ref{RAA_pT_D}, we show $R_{AA}$ of $D$ mesons as a function of $p_T$ for Pb+Pb, Xe+Xe, Ar+Ar and O+O collisions at the LHC energies.
The setup is the same as Fig. \ref{RAA_pT_h}.
Four different centralities (0-10\%, 10-30\%, 30-50\% and 50-80\%) are shown.
One can see that our model can can also describe $D$ meson $R_{AA}$ as a function of $p_T$ and centrality for Pb+Pb collisions as measured by ALICE and CMS Collaborations.
This means that the flavor, energy and system size dependences of jet quenching can be well understood in our jet quenching model.
Using the same setups, the predictions for smaller collision systems (Ar+Ar and O+O collisions) are shown.
We can see that the jet quenching effect on $D$ meson $R_{AA}$ is also quite significant in central O+O collisions.

In Fig. \ref{RAA_Npart_D}, we show $R_{AA}$ of $D$ mesons as a function of $N_{\rm part}$ in four collision systems for two $p_T$ bins: $p_T = (6.7, 7.7)$~GeV in the left panel and $p_T = (26, 30)$~GeV on the right.
Similar to charged hadron $R_{AA}$, $D$ meson $R_{AA}$ is also a decreasing function of $N_{\rm part}$, which clearly shows that jet quenching effect becomes weaker for smaller system sizes.

Figure \ref{RAA_pT_B} shows $B$ meson $R_{AA}$ as a function of $p_T$ for four different collision centralities (0-10\%, 10-30\%, 30-50\% and 50-80\%) in Pb+Pb, Xe+Xe, Ar+Ar and O+O collisions at the LHC energies.
One can see that due to larger masses of $b$ quarks, the values of $B$ meson $R_{AA}$ are larger than $D$ meson and charged hadron $R_{AA}$.
We can see that even for $B$ mesons, jet quenching effect on $R_{AA}$ is still sizable in central O+O collisions.
Figure \ref{RAA_Npart_B} shows $B$ meson $R_{AA}$ as a function of $N_{\rm part}$ in four collision systems for $p_T = (6.7, 7.7)$~GeV (Left) and $p_T = (26, 30)$~GeV (Right).
Similar system size ($N_{\rm part}$) dependence can be seen for $B$ meson $R_{AA}$.
Our results on the flavor, mass, energy and system size dependences of jet quenching can be tested by future measurements.

Recently, experiments have also measured the nuclear modification factor $R_{AA}$ of heavy flavor decayed electrons.
Using the same setups, we further calculate $R_{AA}$ for electrons decayed from $D$ and $B$ mesons for central 0-10\%, 10-30\%, 30-50\% and 50-80\% Pb+Pb, Xe+Xe, Ar+Ar and O+O collisions at the LHC energies, as shown in Fig. \ref{RAA_pT_e}.
One can see that our model can also describe the heavy flavor decayed electron $R_{AA}$ in Pb+Pb collisions at 5.02A~TeV measured by ALICE Collaboration. In addition, we also observe similar system size dependence for heavy flavor decayed electron $R_{AA}$.


\section{Summary}

Using our state-of-the-art jet quenching model, we have performed a systematic study on heavy and light flavor jet quenching in different collision systems at the LHC energies.
The momentum spectra of light and heavy flavor hadrons at high $p_T$ are calculated within a next-to-leading-order perturbative QCD framework, by taking into account both quark and gluon contributions to high $p_T$ light and heavy flavor hadron productions.
The evolution and energy loss of light and heavy partons in the QGP is simulated via the linear Boltzmann transport (LBT) model.
The space-time evolution of the QGP fireball is simulated using a (3+1)-dimensional viscous hydrodynamics.
Using our jet quenching model, we have calculated the nuclear modification factor $R_{AA}$ for charged hadrons, $D$ mesons, $B$ mesons and heavy-flavor decayed electrons as a function of $p_T$ and centrality ($N_{\rm part}$) for central 0-10\%, 10-30\%, 30-50\% and 50-80\% Pb+Pb, Xe+Xe, Ar+Ar and O+O collisions at the LHC energies.
Our model can give a good description of charged hadron, $D$ meson and heavy flavor decayed electron $R_{\rm AA}$'s in central and mid-central Pb+Pb (and Xe+Xe) collisions measured by CMS and ALICE Collaborations.
We have also provided the detailed predictions for the transverse momentum and centrality dependences of $R_{AA}$'s for charged hadrons, $D$ meson, $B$ mesons, and heavy flavor decayed electrons in Pb+Pb, Xe+Xe, Ar+Ar and O+O collisions at the LHC energies.
Our results show that both light and heavy flavor $R_{AA}$'s have a strong system size dependence.
As the system size becomes smaller, jet quenching effect on $R_{AA}$ diminishes: $R_{AA}$ tends to approach unity as the system size shrinks.
Our result indicates that $R_{pA} \approx 1$ in proton-nucleus collisions is mainly due to the small size of the nuclear medium produced in the collisions.
Very interestingly, our calculation also show significant jet quenching effect on $R_{AA}$'s for charged hadrons and heavy flavor hadrons and their decayed electrons in central O+O collisions at the LHC energies, which can be tested by future experiments.
In summary, our study provides an important baseline for studying the flavor, mass, energy and system size dependences of jet quenching.
Further studies along this line should help to identify the unique signatures of mini QGP droplet and search for the disappearance of QGP by scanning the system size and collision energy of relativistic nuclear collisions.


\section{Acknowledgments}

 This work is supported in part by Natural Science Foundation of China (NSFC) under Grants No. 11775095, No. 11890710, No. 11890711 and No. 11935007.  H. X. is supported by the Guangdong Major Project of Basic and Applied Basic Research No. 2020B0301030008, Key Project of Science and Technology of Guangzhou (Grant No. 2019050001), the National Natural Science Foundation of China under Grant No. 12022512, No. 12035007.

\bibliographystyle{plain}
\bibliographystyle{h-physrev5}
\bibliography{refs_GYQ}
\end{document}